\documentclass{emulateapj}

\usepackage{graphicx}
\usepackage{multirow}
\usepackage{xcolor}
\def\be{\begin{equation}}
\def\ee{\end{equation}}

\def\kms{{\rm \,km\,s^{-1}}}

\def\Gyr{{\rm \,Gyr}}
\def\yr{{\rm \,yr}}

\def\Mpc{{\rm \,Mpc}}
\def\kpc{{\rm \,kpc}}
\def\msun{{\,M_\odot}}
\def\hot{{_{\rm hot}}}
\def\calZ{{\cal Z}}

\begin{document}
\title{Probing baryonic processes and gastrophysics in the formation of
the Milky Way dwarf satellites: I. metallicity distribution properties}
\shortauthors{Hou, Yu, \& Lu}

\author{Jun Hou$^1$, Qingjuan Yu$^1$, and Youjun Lu$^2$}
\affil{
$^1$~Kavli Institute for Astronomy and Astrophysics, and School of Physics,
Peking University, Beijing, 100871, China; yuqj@pku.edu.cn \\
$^2$~National Astronomical Observatories, Chinese Academy of
Sciences, Beijing, 100012, China  
}

\begin{abstract}
The Milky Way (MW) dwarf satellites, as the smallest galaxies discovered in the
present-day universe, are potentially powerful probes to various baryonic
processes in galaxy formation occurred in the early universe. In this paper, we
study the chemical properties of the stars in the dwarf satellites around the
MW-like host galaxies, and explore the possible effects of several baryonic
processes, including supernova (SN) feedback, the reionization of the universe
and H$_2$ cooling, on them and how current and future observations may put
some constraints on these processes. We use a semi-analytical model to generate
MW-like galaxies, for which a fiducial model can reproduce the luminosity
function and the stellar metallicity--stellar mass correlation of the MW
dwarfs. Using the simulated MW-like galaxies, we focus on investigating three
metallicity properties of their dwarfs: the stellar metallicity--stellar mass
correlation of the dwarf population, and the metal-poor and metal-rich tails of
the stellar metallicity distribution in individual dwarfs. We find that (1) the
slope of the stellar metallicity--stellar mass correlation is sensitive to the
SN feedback strength and the reionization epoch; (2) the extension of the
metal-rich tails is mainly sensitive to the SN feedback strength; (3) the
extension of the metal-poor tails is mainly sensitive to the reionization
epoch; (4) none of the three chemical properties are sensitive to the H$_2$
cooling process; and (5) comparison of our model results with the current
observational slope of the stellar metallicity--stellar mass relation suggests
that the local universe is reionized earlier than the cosmic average and local
sources may have a significant contribution to the reionization in the local
region, and an intermediate to strong SN feedback strength is preferred.
Future observations of metal-rich and metal-poor tails of stellar metallicity
distributions will put further constraints on the SN feedback and the
reionization processes.
\end{abstract}
 
\section{Introduction}
The dwarf satellite galaxies around the Milky Way (MW), including both
classical dwarf spheroidal galaxies (dSphs) and ultra-faint dwarf galaxies, are
among the least massive galaxies found in the universe
\citep[e.g.,][]{Zucker06, Martin07, Strigari08, Geha09, Simon11,
local_group2012}. They are believed to form at early times of the cosmic
history and reside in small dark matter sub-halos with shallow gravitational
potentials \citep[e.g.,][]{Maccio10, Koposov08, Koposov09}. They are one of the
most representative classes of objects for studying the effects of various
baryonic processes involved in galaxy formation and in the early universe, such
as supernova (SN) feedback \citep[see][]{font2011, Wyithe13, Robertson05},
the reionization of the universe \citep[e.g.,][]{Bullock00, Somerville02,
Benson02,Grebel04, Wyithe06, Bovill09, Munoz09, Busha10, L12}, and molecular
hydrogen cooling (e.g., \citealt{B10}).  Understanding these processes is one
important step to understand the problems that the $\Lambda {\rm CDM}$
cosmology faces at small galactic scales, such as, the `missing satellite
problem' \citep{klypin1999, moore1999, Kravtsov2004, Strigari07, SimonGeha07,
Brooks13} and the `core/cusp problem' (e.g.\ see \citeauthor{core_cusp}
\citeyear{core_cusp} and the references therein). In this paper, we study the
chemical properties of the stars in the dwarf satellites around the MW-like
host galaxies, and explore the possible effects of the above baryonic processes
on them and how current and future observations may put some constraints on
these processes. 

The chemical properties of the dwarfs are potentially powerful probes of the
above baryonic processes (e.g., \citealt{font2011, Frebel10}). The chemical
enrichment of a galaxy are connected directly with the SN feedback process and
the star formation history.  The heavy elements were first synthesized by
nuclear reactions in stars and can be ejected into the interstellar medium at
the later stages of stellar evolution, through stellar winds and SN explosion;
and some chemical-enriched interstellar medium can in turn participate in the
later formation of stars with enhanced metal abundance \citep[see][for a
review]{B10}. The processes may occur repeatedly over time along the star
formation history. Different from the present-day star formation, the star
formation in the early universe can be affected significantly by the
reionization of the universe and the molecular hydrogen cooling processes: the
molecular hydrogen cooling was proposed to be an important cooling mechanism in
forming first stars/galaxies in metal-free mini-halos in the early universe,
where the gas temperature of the mini-halos is not high enough to have
effective atomic cooling (e.g., \citealt{B13}, \citealt{Abel02},
\citealt{Bromm02}); and the reionization of the universe may result in gas
heating-up with increasing pressure and the photoevaporation of small gaseous
halos and hence suppress star formation
\citep[e.g.,][]{gnedin2000,Kravtsov2004,Okamoto08}.

In this paper, we employ the dark matter halo merger trees and the
semi-analytical galaxy formation model \citep[][see also \citealt{WhiteFrenk91,
Kauffmann93, Somerville99}]{galform1} to generate the MW-like galaxies and
their dwarf satellites. The model, with preferred model parameters, can
reproduce the luminosity function of the dwarf satellites, the observed
metallicity versus luminosity/stellar mass correlation of the dwarf population,
and also the stellar metallicity distribution of some individual dwarfs. Armed
with this model, we further explore the effects of various baryonic processes
on the metallicity properties by choosing different recipes for those
processes. By comparing the model results with the observational metallicity
properties of dwarf satellites, we investigate the possibility of using the
metallicity properties to constrain the baryonic processes, such as the
supernova feedback, the reionization of the universe, and the molecular
cooling, involving in the formation processes of those satellites.

The observational metallicity properties have recently been used to understand
the origin of the dwarfs and constrain their formation histories.  For example,
\citet{kirby1,kirby2} use the metallicity distributions of eight MW classical
dSphs to constrain their star formation histories, as well as the chemical
enrichment mode, i.e., the roles of inflow and outflow in the enrichment
histories; and \cite{ferrara2009} use the stellar metallicity--luminosity
correlation and the mean metallicity distribution of ultra-faint dwarfs to
understand their formation sites, redshifts, and star formation histories. Our
study is distinguished from previous works in a few aspects of the purpose and
the method details as follows.
\begin{itemize}
\item We consider the detailed assembling history and the accompanied star
formation history of each dwarf satellite through a semi-analytical modeling
of the MW formation.  Many previous works adopt simple description of the star
formation history for individual dwarfs \citep[e.g.,][]{Carigi02,LM2003,
LM2004, Fenner06, Marcolini06, Marcolini08, kirby2}, which does not include
the detailed assembling history of galaxies and their dark matter halos. 
\item The chemical enrichment caused by SNe II and Ia are different in the time
delay of ejecting enriched materials after the star formation and in the
element abundance of ejected materials \citep[e.g.,][]{MR2001}. We consider the
time delay of the SN Ia enrichment after star formation explicitly in the
chemical enrichment model; while a number of previous works only use simple
instantaneous recycling of ejected materials, which may not address the SN Ia
enrichment properly \citep[e.g.,][]{LM2004}.  We note that the chemical
enrichment of SNe Ia has been considered in some semi-analytical galaxy
formation model (e.g., \citealt{Nagashima1, Nagashima2, Y13}), for example, for
intracluster medium, elliptical galaxies, or MW-like disk galaxies, but not for
galaxies as small as the the dwarfs around the MW. 
\item We quantify the extension of the metal-poor and metal-rich tails in the
stellar metallicity distribution and explore the possibility to use them as a
probe of the baryonic processes. 
\item The semi-analytical galaxy formation models have also been used to
explore some properties of the dwarfs (e.g.,
\citealt{font2011,guo11,L10,RS13}); however, those previous studies focus
mainly on different aspects, e.g., on the number abundance of the dwarfs or the
chemical evolution in an individual (Sculptor) dwarf. \citet{font2011}
illustrate that the chemical properties of the dwarfs can be used to break the
degeneracy in the effects of SN feedback and reionization on their luminosity
function.
\item In addition, we adopt the Monte-Carlo method based on the modified
extended Press-Schechter function to generate dark matter halo merger trees. By
this method, a large number of trees (e.g., 100 or more trees) can be generated
for each set of parameters in an efficient way, which allows a statistical
study of the dwarf chemical properties. Numerical simulations have provided the
assembly histories of several MW-sized halos, e.g., the Via Lactea simulation
\citep{Diemand07,Rocha12} and the Aquarius simulation \citep{Springel08,S13};
however, a much larger number of merger trees for MW-sized halos directly from
numerical simulation are still not available.
\end{itemize}

The paper is organized as follows. The semi-analytical galaxy formation model
used in this study is described briefly in Section~\ref{sec:model}, with
emphases on the recipes of the SN feedback and the SN Ia explosion rate that
implemented. We use the model to generate the MW-like galaxies and their dwarfs
in Section~\ref{sec:result}. The model with a set preferred parameters can
reproduce the observational results on  the metallicity properties of the MW
dwarf satellites well.  We also show the obtained stellar metallicity versus
stellar mass correlation and the metallicity distribution of the dwarfs, by
using different recipes for SN feedback, reionization of the universe, and
molecular hydrogen cooling. Note that in this paper the metallicity is
expressed through [Fe/H], and we do not consider the detailed distribution of
alpha and other elements.  The constraints on the various physical processes
are discussed in Section~\ref{sec:discussion}, and conclusion is given in
Section~\ref{sec:summary}.

We note that the age properties of the dwarf satellites can be also used to put
constraints on the above several baryonic processes, in addition to the
chemical properties investigated in this paper. We shall adopt a similar method
to explore the constraints from the age properties in a subsequent paper (in
preparation).

In this paper we set the Hubble constant as $H_0=100\,h\kms\Mpc$, and the
cosmological model used is $(\Omega_{\rm m},\Omega_\Lambda,h,\sigma_8)=(0.25,0.75,0.70,0.90)$.

\section{Method} \label{sec:model}

In this section, we briefly describe the semi-analytical galaxy formation model
that is used to explore the metallicity properties of the MW satellites.  The
backbone of the model is the merger trees of MW-sized dark matter halos, which
may represent the hierarchical growth history of the MW host halo.  Detailed
semi-analytical recipes for galaxy formation and evolution \citep[for
references, see][]{galform1, WhiteFrenk91, Kauffmann93, Somerville99} are
incorporated into the merger trees to obtain the observational properties of
the central MW galaxy and its satellites.

We plant the merger trees of MW-sized halos by the Monte-Carlo method developed
by \citet[][see also \citealt{Kauffmann93, SK99, galform1}]{merger_tree}, which
is based on a modified version of the extended Press-Schechter formula. The
obtained halo mass functions are in good agreement with the N-body simulation
results.  The merger trees are built from redshift $z=0$ to $20$, with 79 equal
intervals in the logarithm of $1+z$.  The mass resolution of the merger trees
is set to be $1.0 \times 10^6\msun$ , which is the typical mass of mini-halos
that may be important for primordial star formation at high redshift $z \sim
20$.  According to the current constraints on the MW halo mass (e.g.,
\citealt{BK13} and references therein), i.e., $1-2\times 10^{12}\msun$, we
obtain the merger trees for halos with the present-day mass $M_{\rm halo}$ of
$1.0 \times 10^{12}$ and $2.0 \times 10^{12}\msun$, respectively. 

The semi-analytical galaxy formation model in this study is based on GALFORM
\citep{galform1, galform2,galform3}, but with several modifications.  As
demonstrated by \citet{galform1}, \citet{Kauffmann93}, \citet{Somerville99},
\citet{Croton06}, and \citet{galform3}, the semi-analytical models can
successfully reproduce a number of observations on the statistical
distributions of galaxy properties, including the galaxy luminosity function,
the stellar mass function, etc. Some constraints on the parameters involving
in the semi-analytical model have been obtained by using those observations;
however, there are still large degeneracies among those parameters as
suggested by recent studies exploring the parameter space of the
semi-analytical galaxy formation model \citep[e.g.,][]{LuMo11, LuMo12,
Bower2010}.  The metallicity properties studied in our study may also put some
constraints on the model parameters, especially those characterized SN
feedback, the reionization of the universe, molecular cooling in the early
universe. We vary the model parameters involving in these several processes to
check the effects of those parameters on the metallicity properties of the MW
dwarfs. 

\begin{figure} \center \includegraphics[scale=0.6]{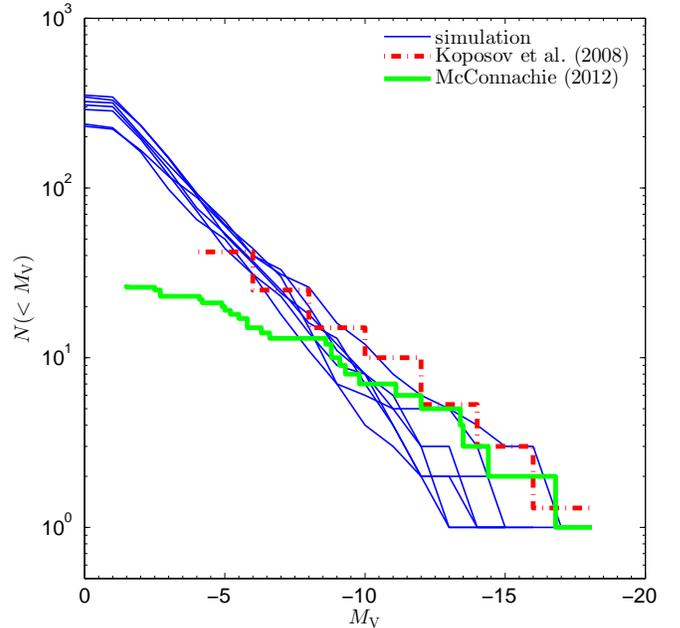}
\caption{An example of the cumulative luminosity functions of the dwarf
satellites of the MW-like galaxies obtained from our models. The $y$-axis
represent the number of the satellites with absolute magnitude brighter than
$M_V$. The seven blue solid lines are the simulation results of the first
parameter set listed in Table~\ref{tab:parameter_sets}, and each line is for
one MW-like host galaxy.  The green line shows the observational result for 27
MW satellites \citep{local_group2012}, and the red one is the result after
doing the searching volume correction (\citealt{Koposov08}; see also
\citealt{Tollerud08}).  Our simulation results generally reproduce the
observational luminosity function of the MW dwarf satellites.
}\label{fig:LF} \end{figure}

The related recipes, e.g., on SN feedback, reionization, and molecular hydrogen
cooling, are summarized below.  For each set of the recipe parameters, we
generate $100$ merger trees and apply the semi-analytical recipes to them.  The
various related baryonic processes (e.g., star formation, chemical enrichment,
and galaxy mergers) are calculated in a time step $\Delta t \sim 10^6 {\rm
yr}$.  From the obtained present-day galaxies, we select the MW-like hosts for
the study of their satellites. The MW-like hosts are selected by the criteria
that the total stellar mass of a present-day host galaxy is in the range of
$4\-- 6 \times 10^{10}\msun$ and its bulge mass-to-disk mass ratio is between
$0.1$ and $0.4$.  Along the assembling history of the MW-sized halo, a
satellite of the present-day MW galaxy was a host galaxy of a small halo at an
early time before it fell into a big halo. An example of the dwarf
satellite luminosity function obtained from our models is illustrated in
Figure~\ref{fig:LF}.

As seen from Table~\ref{tab:parameter_sets} to be listed in
Section~\ref{sec:result} below, for some recipe parameter sets, the models fail
to generate the MW-like host galaxy. In these cases, we just choose randomly
$10$ results of the corresponding parameter set to illustrate the effects of
the baryonic processes. 
\begin{itemize}
\item {\it The reionization of the universe:} the reionization in the early
universe reduces the baryon fraction of a dark matter halo. During the
reionization epoch, the intergalactic medium (IGM) heats up with increasing
pressure, which can suppress the collapse of the IGM onto dark matter halos,
and the gas that was previously in a dark matter halo can also evaporate out of
the halo. The extent of the reduction in the baryon fraction of a dark matter
halo depends on the halo gravitational potential or roughly the halo mass.  The
dependence is modeled through a mass scale called the `filtering mass' ($M_{\rm
F}$) as follows,
\begin{equation}
f_{\rm b}=\frac{1}{[1+(2^{1/3}-1)M_{\rm F}/M_{\rm halo}]^3}
\end{equation}
where $f_{\rm b}$ is the ratio of the baryon fraction in a halo with mass
$M_{\rm halo}$ to the cosmic average baryon fraction, and $M_{\rm F}$ is a
function of redshift, as well as a function of the completion redshift and the
duration of the reionization process.  A halo with mass $M_{\rm halo}<M_{\rm
F}$ loses more than 50\% of baryonic matter expected by the cosmic average.  In
this paper, we use Equations (B1) and (B2) in \cite{Kravtsov2004} (see also
\citealt{gnedin2000, Okamoto08}) to calculate $M_{\rm F}$ and model the effects of the
reionization.  This reionization recipe for $M_{\rm F}$ is characterized by two
parameters: $z_0$, the redshift that the first ionized bubble formed, and
$z_{\rm r}$, the completion redshift of the reionization.

Two sets of the reionization parameters are used in our model, i.e.,
$(z_0,z_{\rm r})=(15, 10)$ and $(10, 7)$.  Observations on the highest redshift
QSOs suggest that the reionization process is completed at redshift $z\sim 6-7$
(e.g., \citealt{Becker01,Mortlock11}), and the polarized cosmic microwave
background radiation detected by WMAP and PLANCK suggests that the universe
probably began to be reionized at redshift $z \ga 10$
\citep[e.g.,][]{WMAP9,PLANCK1}. According to these observations, the
reionization model set by $(z_0, z_{\rm r}) = (10, 7)$, suggested by
\citet{Kravtsov2004}, may represent a case close to or slightly later than the
cosmic average reionization epoch of the real universe, and the other model set
by $(z_0, z_{\rm r})= (15, 10)$ may represent a case a little earlier than
the cosmic average of the real universe.  Hereafter, we refer the former case
as `weak reionization' (or `late reionization') and the latter case as `strong
reionization' (or `early reionization'). Note that \citet{font2011} also adopt
$(z_0, z_{\rm r})=(15, 10)$ as a `strong reionization' scenario and suggest
that the contribution from local sources may lead to an earlier reionization of
the local patch compared with the cosmic average.

\item {\it Molecular hydrogen cooling:} we model the molecular hydrogen cooling
process occurred in the early universe by using Equations (21)--(29) in
\cite{B10} (see also \citealt{galli1998}) and we do not allow any molecular
hydrogen cooling after the completeness of the reionization, because the
abundance of hydrogen molecules would be significantly suppressed by the strong
UV background.
 
\item {\it Cooling of hot halo gas:} The gas cooling recipe of the GALFORM is
modified in this work.  The cooling recipe in \citet{galform1} is likely to
underestimate the amount of cooling gas in a halo, where the gas reheated by
SNe is assumed not to participate into the cooling process until the dark
matter halo doubles its mass along the merger trees.  The cooling recipe in
\citet{galform3} improves this but possibly overestimates the amount of 
cooling, and this model has to adopt an extremely strong SN feedback (e.g.,
$v\hot=485\kms$ and $\alpha\hot=3.2$ in Eq.~\ref{eq:scale_law2} below) to
balance the overestimated cooling amount and reproduce the observational galaxy
luminosity function. \citet{Benson2010} further modify the cooling recipe by
continuously updating estimate of cooling time and halo properties at each
timestep. The model result of this update can match the observational galaxy
luminosity function, but the gas phase metallicity is underestimated for 
relatively low-luminosity galaxies, where SN feedback is argued to be the main
driver for the relation between the galaxy luminosity and the gas phase
metallicity and the adopted strength is still high (with $v\hot\sim328$ or
$358\kms$ and $\alpha\hot=3.36$; see Table 5 therein); and thus the cooling
rate is likely to be still overestimated.  In this work, we employ a cooling
recipe somewhat between the recipes in \citet{galform1} and \citet{galform3}.
This modified recipe is detailed in the Appendix.
 
\item {\it SN feedback efficiency:} SN feedback reheats the cold gas in a
galaxy, expels it out of the galaxy, and enriches the metallicity in the halo
gas.  Following \cite{galform1}, we use the formula
\begin{eqnarray}
  & & dM_{\rm reheat}\ =\ \beta \psi dt \label{eq:scale_law1}\\
  & & \beta\ =\ (v_{\rm disk}/v_{\rm hot})^{-\alpha_{\rm hot}}
\label{eq:scale_law2}
\end{eqnarray}
to estimate the amount of cold gas expelled from a galaxy by SNe II, where
$dM_{\rm reheat}$ is the mass of the gas reheated by SN feedback during time
interval $dt$, $\psi$ is the star formation rate, $\beta$ is the SN feedback
efficiency, $v_{\rm disk}$ is the circular velocity of the galaxy disk, and
$\alpha_{\rm hot}$ and $v_{\rm hot}$ are two parameters defining the strength
of the feedback.  We call Equation (\ref{eq:scale_law2}) the SN feedback
efficiency scaling law. 

Once the SN explosion energy is sufficiently high, all the reheated gas may be
expelled out of the galaxy if the following energy condition is satisfied,
i.e.,
\be
dE_{\rm SN}-\frac{1}{2}v_{\rm vir}^2 dM_{\rm reheat}\ge 0,
\label{eq:energy_condition}
\ee
where $dE_{\rm SN}=\epsilon_{\rm halo}\times \frac{1}{2}v_{\rm SN}^2\psi dt$ is
the total energy released by SNe II and coupling to the IGM during time $dt$,
$v_{\rm vir}$ is the virial velocity of the halo, $\frac{1}{2}v_{\rm SN}^2$ is
the total energy released per unit mass by SNe with $v_{\rm SN}=630\kms$ for
the Chabrier IMF \citep{chabrier2003}, and $\epsilon_{\rm halo}=0.05$ is the
fraction of the energy that couples to the cold gas in the disk (e.g.,
\citealt{L10}).  If the energy condition (Inequality \ref{eq:energy_condition})
is satisfied, the reheated gas can either be ejected into the dark matter halo
or even escape out of the halo, with masses approximated by 
\begin{eqnarray}
  &  & dM_{\rm outflow}   \nonumber \\
  & = & \min\left(\frac{dE_{\rm SN}-\frac{1}{2}v_{\rm vir}^2 dM_{\rm
reheat}}{\frac{1}{2}v_{\rm vir}^2},dM_{\rm reheat}\right), \label{eq:dMoutflow}
\end{eqnarray} and
\be
  dM_{\rm stay}  =  dM_{\rm reheat}-dM_{\rm outflow},  \label{eq:dMstay}
\ee
where $dM_{\rm outflow}$ is the mass of the gas running out of the halo during
time $dt$ and $dM_{\rm stay}$ is the mass of the reheated gas staying in the
halo. We assume that the outflow returns to the halo on a halo dynamical time
scale ($\tau_{\rm halo}\equiv r_{\rm vir}/v_{\rm vir}$, where $r_{\rm vir}$ is
the halo virial radius) as 
\begin{equation}
dM_{\rm back}=\frac{M_{\rm outflow}}{\tau_{\rm halo}}dt.
\end{equation}
If the supernova explosion is energetic enough to expel all the reheated gas
out of the dark matter halo, i.e., $(dE_{\rm SN}-\frac{1}{2}v_{\rm vir}^2
dM_{\rm reheat})/(\frac{1}{2}v_{\rm vir}^2)>dM_{\rm reheat}$ in Equation
(\ref{eq:dMoutflow}), Equations (\ref{eq:dMoutflow}) and (\ref{eq:dMstay})
reduce to
\begin{eqnarray}
& & dM_{\rm outflow}=dM_{\rm reheat}, \label{eq:dMoutflowall}\\
& & dM_{\rm stay}=0. \label{eq:dMstay0}
\end{eqnarray}

If the energy condition is not satisfied, we use
\begin{eqnarray}
& & dM_{\rm outflow}=0, \label{eq:energy_condition1}\\
& & dM_{\rm stay}=dE_{\rm SN}/(\frac{1}{2}v_{\rm vir}^2),
\label{eq:energy_condition2}
\end{eqnarray}
to determine the masses of the gas that goes out of the halo and stays in the
halo, which is equivalent to replace the SN feedback scaling law with
\begin{eqnarray}
\beta =\beta_E & \equiv & dE_{\rm SN}/(\frac{1}{2}v_{\rm vir}^2\psi dt) =\epsilon_{\rm
halo}(v_{\rm vir}/v_{\rm SN})^{-2} 
\label{eq:betaE}
\end{eqnarray}
in Equation (\ref{eq:scale_law1}) above. Equation~(\ref{eq:betaE}) represents
the supernova feedback efficiency limit set by the energy condition.

Figure~\ref{fig:beta_figure} shows the relation between the SN feedback
efficiency given by the scaling law and the limit set by the energy condition.
The solid lines with different symbols show the feedback efficiency given by
the scaling law with different values of $v_{\rm hot}$ and $\alpha_{\rm hot}$.
Specifically, the solid line without symbols indicates the efficiency for the
fiducial model ($v_{\rm hot}=200\kms$ and $\alpha_{\rm hot}$=3.2, see the
fiducial model defined below in Section~\ref{sec:result}). The dashed line
gives the SN feedback efficiency limit set by the energy condition (Inequality
\ref{eq:energy_condition}); and the feedback efficiency is allowed by the
energy condition in the region below the dashed line, while not allowed in the
shaded region above. In the shaded region, Equations
(\ref{eq:energy_condition1}) and (\ref{eq:energy_condition2}) are applied.  The
dotted line indicates the efficiency below which all the reheated gas is
ejected out of the halo, where Equations (\ref{eq:dMoutflowall}) and
(\ref{eq:dMstay0}) are applied. The dashed and the dotted lines are drawn under
the assumption $v_{\rm vir}=v_{\rm disk}$, and the deviation from the
assumption is somewhat significant only at low velocities. The region between
the dotted line and the dashed line have $dM_{\rm outflow}>0$ and $dM_{\rm
stay}>0$, described by Equations (\ref{eq:dMoutflow}) and (\ref{eq:dMstay}).

\begin{figure} \center
\includegraphics[scale=0.5]{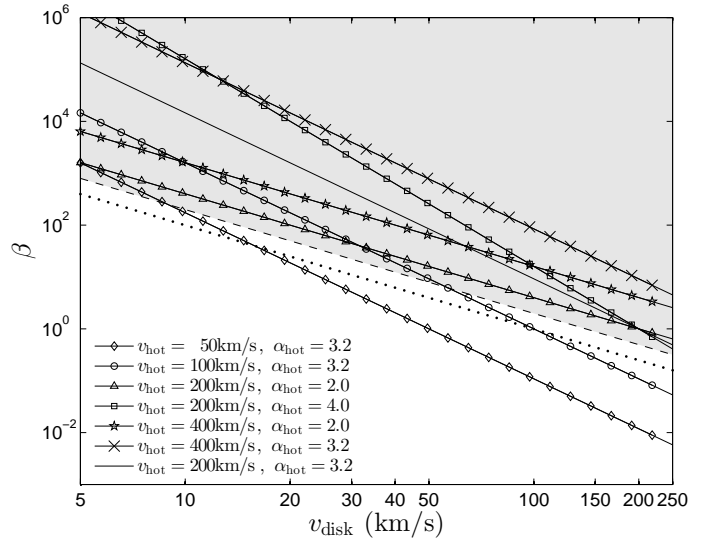}
\caption{The SN feedback efficiency $\beta$ given by the feedback scaling law
as a function of $v_{\rm disk}$ and the limit set by the energy condition. The
solid lines represent the feedback scaling laws with different values of
$v_{\rm hot}$ and $\alpha_{\rm hot}$, and specifically, the solid line without
symbols is for the fiducial model with $v_{\rm hot}=200\kms$ and $\alpha_{\rm
hot}=3.2$.  The dashed line represents the limit set by the energy condition.
The feedback efficiency $\beta$ (Eq.~\ref{eq:scale_law2}) is allowed by the
energy condition in the region below the dashed line, while not allowed in the
gray shaded region above it. The dotted line indicates the feedback efficiency
below which all the reheated gas would be ejected out of the halo.  The dashed
and the dotted lines are drawn under the assumption $v_{\rm disk}=v_{\rm vir}$,
which is a plausible assumption in general; and the efficiency limits may
scatter around the lines due to the scatter of $v_{\rm vir}$ around $v_{\rm
disk}$.
}\label{fig:beta_figure} \end{figure}

As mentioned before, along the assembling history of the MW-like halo, a
satellite of the present-day MW may be a host galaxy of a small isolated halo
before it falls into a big halo at an early time. We apply the above energy
condition only to a galaxy before it becomes a satellite. We do not apply it to
satellites, but assume that the reheated gas from satellites (with mass
expected by Eqs.~\ref{eq:scale_law1} and \ref{eq:scale_law2}) is expelled into
the big host halo, as the original halos of the satellites are largely tidally
disrupted along their motion in the big host halo, and the tidal field induced
by the big host halo also helps to keep those expelled materials out of the
satellites. 

\item {\it Metallicity production:} In this work, the Fe yield of SNe II is
adopted from tables 2--3 in \citet{SN_II_pattern}, and the Fe yield of SNe Ia
is from \citet{SN_Ia_pattern}.  The metals ejected by SNe are assumed to be
homogeneously and instantaneously mixed with the interstellar medium in the
galaxy; and after the mixture, some metals can be ejected out of the galaxy
along with the mixed interstellar medium that is ejected out by SN explosions.

\item {\it Chemical enrichment due to SN Ia explosions:} The feedback due to SN
Ia explosions is explicitly included in this work.  We assume that the energy
released by a SN II explosion and a SN Ia explosion is the same. We use the
same feedback recipe for them
(Eqs.~\ref{eq:scale_law1}--\ref{eq:energy_condition2}), but include the
non-negligible time delay between the formation of SN Ia progenitors and the SN
Ia explosions.  In the model, the number of SN Ia explosions within a time
interval $\Delta t$ which begins at a given time $t$, denoted by $N_{\rm
Ia}(t)$, is given by
\begin{eqnarray}
N_{\rm Ia}(t) & = & \int_t^{t+\Delta t}dt'' \int_0^{t''} R_{\rm
Ia}(t''-t')\psi(t')dt', \nonumber \\
 & \simeq & \Delta t\times\int_0^t R_{\rm Ia}(t-t')\psi(t')dt',
\label{eq:NIa}
\end{eqnarray}
where $t''-t'$ is the stellar population age, $R_{\rm Ia}$ is the SN Ia
explosion rate as a function of the time delay and in units of number per time
per mass (e.g., $\Gyr^{-1}\msun^{-1}$), and $\Delta t\sim 10^6\yr$ is small
enough compared to the time delay of SNe Ia explosion assumed below.  According
to the observational SN Ia rate $R_{\rm Ia}$ reported by \cite{maoz2010}, a
time delay of $0.1\Gyr$ is assumed to generate SN Ia since the formation of a
stellar population, and the number of the generated SNe decreases with
increasing age of this population as a power law.  \cite{Nagashima1,
Nagashima2} also considered the chemical enrichment due to SN Ia explosions in
their semi-analytical galaxy formation model, where the star formation history
of a galaxy is re-binned into $30$ rough time bins to obtain the convolution
integral of Equation (\ref{eq:NIa}).  In this work, we still use the star
formation history obtained with small timesteps $\Delta t$, but approximate the
power-law decline of the SN Ia explosion rate as a combination of the linear
functions and the exponential function, which can expedite the calculation of
the convolution integral and obtain $N_{\rm Ia}(t)$ recursively and efficiently
in each timestep. The method to calculate the SN Ia explosion rate is detailed
below. 

We approximate the power-law rate into $6$ continuous linear segments when the
stellar population age $\tau$ is between $0.1$ and $2\Gyr$. The $i$-th segment
approximates the power-law rate when $\tau$ is from $\tau_{i-1}$ to $\tau_i$
($i=1,2,...,6$), with $\tau_0=0.1\Gyr$ and $\tau_6=2\Gyr$.  When the stellar
population age is older than $2\Gyr$, we approximate the power-law rate by an
exponential tail $C_{\rm Ia}\exp (c_0t')$, where $C_{\rm Ia}$ is a
normalization factor and $c_0$ is a parameter, and we adopt $C_{\rm Ia}=5\times
10^{-4} \Gyr^{-1}\msun^{-1}$.  Note that \cite{maoz2010} report $C_{\rm
Ia}=1\times 10^{-3}\Gyr^{-1}\msun^{-1}$, and \cite{kirby2} mention that the
observations in \cite{maoz2010} can easily be consistent with half of that
value. The parameter $c_{0}$ is fixed by requiring the exponential tail and the
sixth segment give the same value at $\tau=2 {\rm Gyr}$. The relative error in
$R_{\rm Ia}$ caused by the approximation described above is about 0.15\%. With
the approximation, we have the following recursive formulas for $N_{\rm Ia}$,
with which the computational complexity is significantly reduced,
\begin{eqnarray}
N_{\rm Ia}(t+\Delta t) & = & N_{\rm Ia}(t)+ \sum_{i=1}^{6} k_{i}\Delta t^2
M_{i}(t+\Delta t) \nonumber \\ & + &  (k_{1}\tau_0+b_{1})\psi (t-\tau_0+\Delta
t)\Delta t^2 \nonumber \\
	& + &  N_{\rm exp}(t+\Delta t)-N_{\rm exp}(t), \nonumber \\
\end{eqnarray} where $k_{i}$ and $b_{i}$ are the slope and intercept of the
$i$-th linear segment of the approximated SN Ia explosion rate, respectively,
and
\begin{eqnarray}
M_{i}(t+\Delta t) & = & M_{i}(t)-\psi(t-\tau_i+\Delta t)\Delta t \nonumber \\ &
& +  \psi(t-\tau_{i-1}+\Delta t)\Delta t, \\
N_{\rm exp}(t+\Delta t) & = & N_{\rm exp}(t)\exp (c_{0}\Delta t) \nonumber \\ &
&   +  \psi(t-\tau_6+\Delta t)C_{\rm Ia}\exp (c_{0}\tau_6) \Delta t^2,
\nonumber \\ \\
M_{i}(0) & = & 0, \nonumber \\ N_{\rm exp}(0) & = & 0, \nonumber 
\end{eqnarray}
are used to define and calculate $M_{i}$ ($i=1,2,...,6$) and $N_{\rm exp}$.
\end{itemize}

Note that tidal disruption of the original halo of a satellite after its
infalling into a big host halo is included when considering the effect of SN
feedback, as mentioned above; but we ignore the tidal stripping and disruption
of its stellar and cold gas components in our model. Compared with the original
halo size of the satellite, the stellar and cold gas components are located in
a smaller central region of the halo, which should be affected less by tidal
effects from the big host halo. In addition, \citet{S13} show that the tidal
stripping and disruption of satellites have a small effect on the satellite
total luminosity function. 

\section{Results} \label{sec:result}
In this section, we try different recipe parameters for the processes of the
supernova feedback, the reionization of the universe, and the molecular
hydrogen cooling. The different sets of the parameters are listed in
Table~\ref{tab:parameter_sets}.  We find one set of parameters that can
reproduce some observational properties of the MW dwarfs better than the
others, including the satellite luminosity function, their luminosity/stellar
mass versus stellar metallicity correlation, and the metallicity distributions
of the classical dSphs, as well as the host galaxy properties (stellar mass,
bulge-to-disk mass ratio); and we denote the model with this set of parameters
by `the fiducial model' and list it as the first parameter set in
Table~\ref{tab:parameter_sets}.  Comparison of the results obtained with the
other different parameter sets helps us to investigate the effects of the
physical processes. We present our model results on the stellar mass versus
stellar metallicity correlations of the dwarfs in Section~\ref{sec:sub1} and
the stellar metallicity distribution in individual dwarfs in
Section~\ref{sec:z_distribution}.

\begin{table*}[ht]
\caption{The parameter sets used in the paper}
\begin{center}
\begin{tabular}{c|c|c|c|c|c|c} 
\hline
\hline
$v_{\rm hot} ({\kms})$ & $\alpha_{\rm hot}$ & $z_0$ & $z_{\rm r}$ & $M_{\rm halo} (\msun)$ & ${\rm H}_2$ cooling & MW-like galaxies\\
\hline
        $200$          &      $3.2$     &  $15$ &  $10$ & $2\times 10^{12}$ & on & $7$ \\
\hline
        $400$          &      $3.2$     &  $15$ &  $10$ &\multirow{7}{*}{$2\times 10^{12}$}& on & 0       \\
        $100$          &      $3.2$     &  $15$ &  $10$ &   & on & 0       \\
         $50$          &      $3.2$     &  $15$ &  $10$ &   & on & 0       \\
        $200$          &      $4.0$     &  $15$ &  $10$ &   & on &  $9$     \\
        $200$          &      $2.0$     &  $15$ &  $10$ &   & on & $18$     \\
        $200$          &      $3.2$     &  $10$ &   $7$ &   & on &  $8$     \\
        $200$          &      $3.2$     &  $15$ &  $10$ &   & off&  $9$     \\
\hline
        $400$          &      $3.2$     &  $15$ &  $10$ &\multirow{7}{*}{$1\times 10^{12}$}& on & 0       \\
        $100$          &      $3.2$     &  $15$ &  $10$ &   & on & $12$      \\
         $50$          &      $3.2$     &  $15$ &  $10$ &   & on & $13$       \\
        $200$          &      $4.0$     &  $15$ &  $10$ &   & on &  $9$     \\
        $200$          &      $2.0$     &  $15$ &  $10$ &   & on & $10$     \\
        $200$          &      $3.2$     &  $10$ &   $7$ &   & on & $17$     \\
        $200$          &      $3.2$     &  $15$ &  $10$ &   & off&  $7$     \\
\hline
\end{tabular}
\end{center}
\tablecomments{ The first parameter set
is for the fiducial model. The $v_{\rm hot}$ and $\alpha_{\rm
hot}$ are the parameters in the feedback scaling law, $z_0$ and $z_r$ represent
the beginning and completion redshifts of the reionization, $M_{\rm halo}$ is
the present-day halo mass, the ``on'' and ``off'' represent whether the process
of the molecular hydrogen cooling in the early universe is switched on or not. For each
parameter set, we construct 100 merger trees and obtain 100 central galaxies, and the last
column gives the number of the central galaxies that are MW-like.
\label{tab:parameter_sets}}
\end{table*}

We find that the chemical properties of the satellites obtained from $M_{\rm
halo}=1\times 10^{12}$ and $2\times 10^{12}\msun$ do not differ much, so below
we only show the results of $M_{\rm halo}=2\times 10^{12}\msun$ for brevity.

\subsection{The stellar mass -- metallicity correlation} \label{sec:sub1}

\begin{figure*} \center
\includegraphics[scale=1.0]{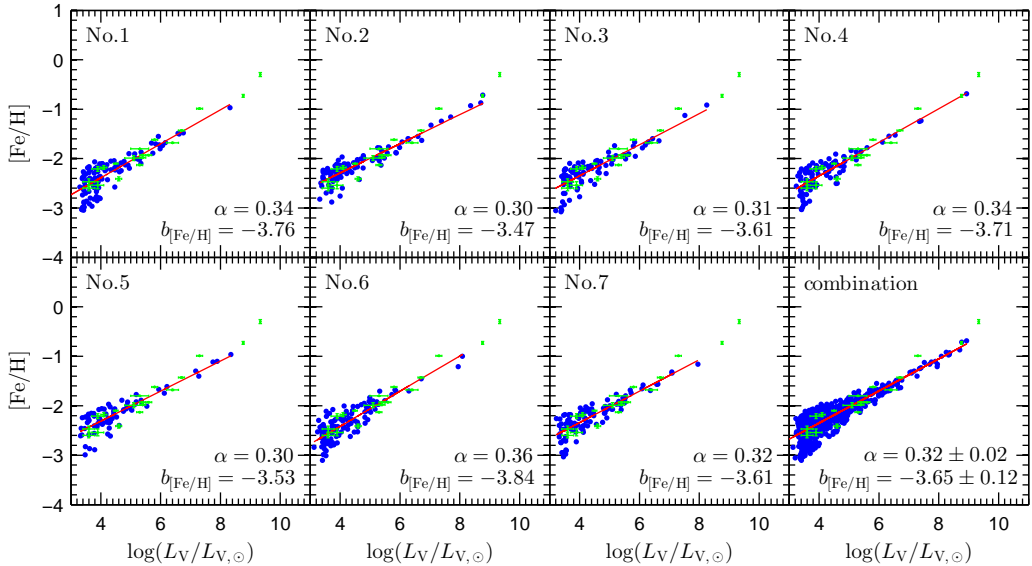}
\caption{The stellar metallicity -- luminosity correlations of the satellites
of the MW-like galaxies. In the first seven panels, the blue dots represent the
satellites from the simulations. The green dots and their error bars are the
data from the observations (\citeauthor{kirby1} \citeyear{kirby1} for Fornax,
Leo I, Sculptor, Leo II, Sextans, Draco, CVn I, Ursa Minor, Leo T, Hercules,
Ursa Major I, Leo IV, CVn II, Ursa Major II, ComB, and \citeauthor{martin2008}
\citeyear{martin2008} for Bootes I, \citeauthor{helmi2006} \citeyear{helmi2006}
for Carina, and \citeauthor{vdb2000} \citeyear{vdb2000} for the Large
Magellanic Cloud and the Small Magellanic Cloud). In the first seven panels,
the red lines are the best linear fits to the simulations, where $\alpha$ is
the best fit slope and $b_{\rm [Fe/H]}$ is the best-fit intercept at
$\log(L_{\rm V}/L_{{\rm V},\odot})=0$; and that in the last panel shows the
statistical average of the results shown in the first seven panels. The
parameters labeled in the last panel represent the statistical mean and
standard deviation of the seven best-fit slopes and intercepts.
\label{fig:l_z_correlation}}
\end{figure*}

Figure~\ref{fig:l_z_correlation} shows the stellar metallicity versus
luminosity correlation of the dwarfs generated by the fiducial model, as well
as the observational results \citep{kirby1,martin2008,helmi2006,vdb2000}. As
seen from Table~\ref{tab:parameter_sets}, 7 individual simulations can generate
a MW-like host galaxy, and the first $7$ panels in
Figure~\ref{fig:l_z_correlation} show each result of the individual
simulations. The last panel shows the results of these simulations together. In
each of the first seven panels, the red line shows the best fit to the
simulation results by using the least squares method; $\alpha$ is the best-fit
slope, and $b_{\rm [Fe/H]}$ is the best-fit intercept at $\log(L_{\rm
V}/L_{{\rm V},\odot})=0$.  The red line in the last panel shows the statistical
average of the best-fits in the first 7 panels, and the parameters labeled
represent the statistical mean and standard deviation of the 7 best-fit slopes
and intercepts.
The fiducial model reproduces the observations well.  While the stellar
metallicity -- luminosity correlation is convenient for comparison with
observations,
in this work we use the stellar mass -- metallicity correlation of the dwarfs
obtained from simulations to study the effects of supernova feedback,
reionization, and molecular hydrogen cooling, as the physical processes affect
the star formation histories of the dwarfs directly.  The faintest satellite in
the observation sample shown in Figure~\ref{fig:l_z_correlation} has
$\log(L_{\rm V}/L_{{\rm V},\odot})=3.6\pm 0.2$, and in our simulations the
satellites with this luminosity roughly have stellar masses $\sim 10^4\msun$.
Thus in our study of the stellar mass -- metallicity correlations below, we
adopt $10^4\msun$ as the cut-off mass at the low-mass end of the satellites.

We show the dependence of the stellar mass -- metallicity correlation on the
SN feedback parameters, the reionization, and molecular hydrogen cooling below
in Sections~\ref{sec:depvhot}--\ref{sec:depH2cooling}.

\subsubsection{Dependence on SN feedback}
\label{sec:depvhot}

We study the dependence of the stellar mass -- stellar metallicity correlation
on SN feedback through its dependence on the parameters ($v_{\rm
hot},\alpha_{\rm hot}$).

\begin{figure*} \center
\includegraphics[scale=1.0]{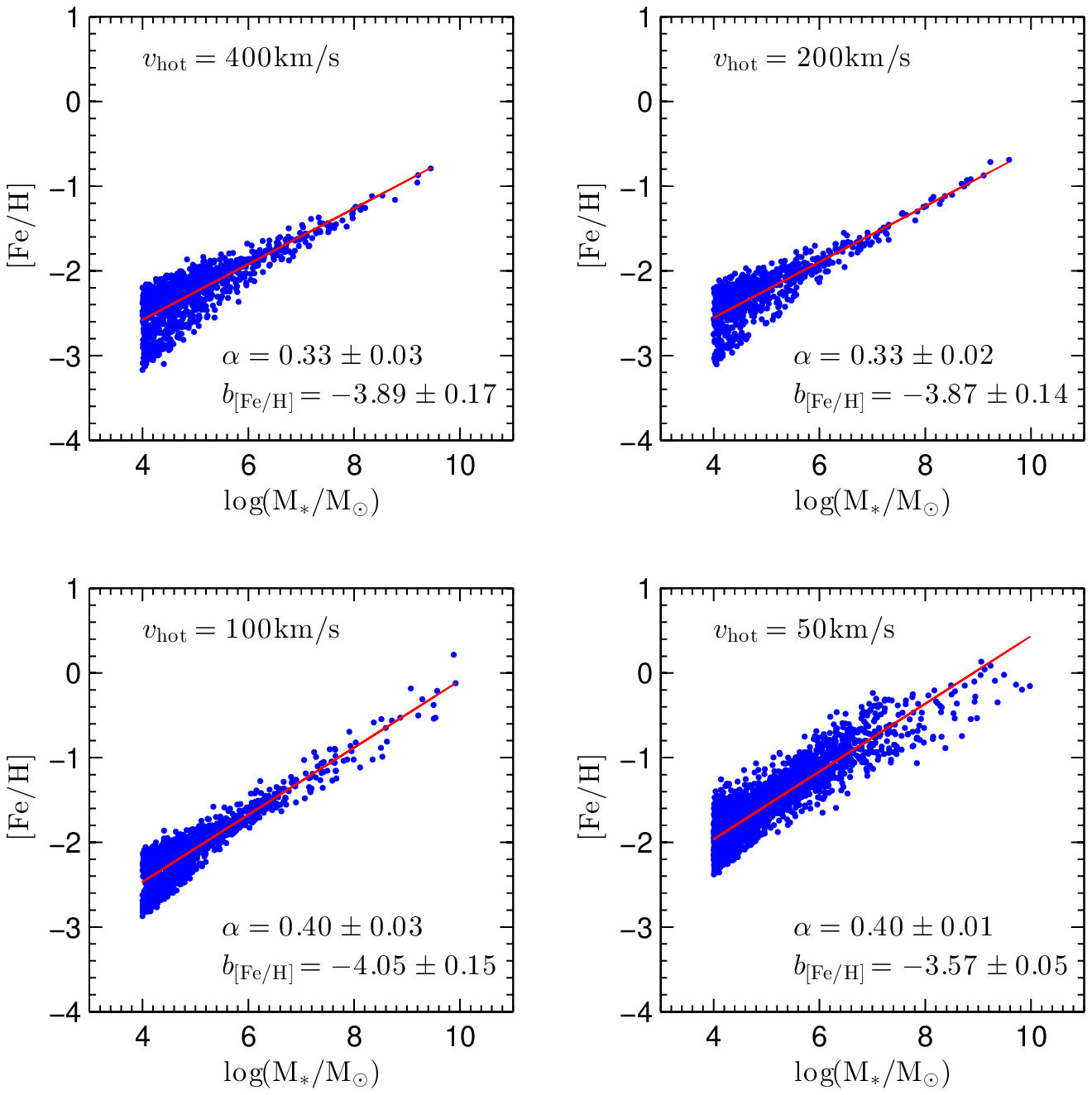}
\caption{The stellar mass -- metallicity correlations obtained from models with
different $v_{\rm hot}$. The blue dots are the simulation results and the red
lines are the linear fits. The value of $v_{\rm hot}$, the mean and the
standard deviation of the slope of the linear fit $\alpha$ and its intercept
$b_{\rm [Fe/H]}$ at $\log(M_*/\msun)=0$ are shown in each panel. The parameters
different from $v_{\rm hot}$ all have the same values as those for the fiducial
model.  As seen from the figure, the distribution range of the blue dots are
almost the same when $v_{\rm hot}>100\kms$ (top panels), as the SN feedback
efficiency in these models is limited by the energy condition (see
Eq.~\ref{eq:energy_condition} and Fig.~\ref{fig:beta_figure}); and it becomes
different for smaller $v_{\rm hot}$ with lower SN feedback efficiency (bottom
panels). The scatter in the star formation durations and the chemical
enrichment caused by SNe Ia contribute significantly to the large scatter of
the correlations at the low-mass end. See details in Section~\ref{sec:depvhot}.
\label{fig:m_z_cor_vhot}}
\end{figure*}

Figure~\ref{fig:m_z_cor_vhot} shows the stellar mass -- stellar metallicity
correlations obtained from the models with different values of $v_{\rm hot}$ in
different panels. Each panel shows the combined results of the corresponding
simulations together, as the last panel of Figure \ref{fig:l_z_correlation}
does. As seen from Figure~\ref{fig:m_z_cor_vhot}, the models with $v_{\rm
hot}=200\kms$ and $v_{\rm hot}=400\kms$ provide almost the same slope. The
model with $v_{\rm hot}=100\kms$ gives a steeper slope, but the metallicities
of the low-mass systems (with satellites mass about several times $10^4\msun$)
are almost not changed. The result of the model with $v_{\rm hot}=50\kms$
cannot be fit well with a single power law, and a turn-off of the correlation
appears at about $10^7\msun$; and the metallicities of the satellites in the
whole mass range, i.e.\ from $10^4\msun$ to $10^{10}\msun$, are all higher than
those predicted by the fiducial model. 

The above dependence on $v_{\rm hot}$ can be understood as the results of the
combination of the feedback efficiency scaling with $v_{\rm hot}$ as a power
law (Eqs.~\ref{eq:scale_law1} and \ref{eq:scale_law2}) and the energy condition
that is expressed by Inequality (\ref{eq:energy_condition}), which can be seen
from Figure~\ref{fig:beta_figure}.
\begin{itemize}
\item In the models with $v_{\rm hot}=200\kms$ and $400\kms$,
as seen from Figure \ref{fig:beta_figure}, the lines for the fiducial model
(solid line) and for the model with $v_{\rm hot}=400\kms$ (with cross symbols)
are all in the gray shaded region, which means the feedback efficiency given by
the feedback scaling laws is too strong to be allowed by the energy condition. In
these systems the effects of the feedback are determined by Equations
(\ref{eq:energy_condition1}) and (\ref{eq:energy_condition2}), independent of
the exact value of $v_{\rm hot}$; so these two models predict almost the same
correlation. Due to the strangulation of the hot gas in satellites and the
strong feedback used in these models, the correlation is generally shaped
before the galaxies become satellites.
\item In the model with $v_{\rm hot}=100\kms$, 
this feedback efficiency given by the feedback scaling law is allowed by the
energy condition in the relatively massive systems, but forbidden in the least
massive system with masses about $10^4\msun$.  As can be seen from
Figure~\ref{fig:beta_figure}, the line for $v_{\rm hot}=100\kms$ (with circle
symbols) is below the shaded region at $v_{\rm disk}\ga 50\kms$, while it is
located within the shaded region at smaller $v_{\rm disk}$.
In the least massive systems located within the shaded region, the feedback is
still determined by the energy condition, and the effects of the feedback is
the same as that in the models with $v_{\rm hot}=200\kms$ and $400\kms$, and so
their metallicities are fairly not changed.  But for the relatively massive
systems located below the shaded region, the feedback efficiency is smaller
than that given by the dashed line and thus results in more efficient metal
enrichment in the disks, which increases the slope of the correlation.
\item In the model with $v_{\rm hot}=50\kms$, the feedback efficiency predicted
by the scaling law is much smaller than that in the fiducial model, and it is
allowed by the energy condition in most of the mass range considered here
($10^4 \-- 10^{10}\msun$). As shown in Figure~\ref{fig:beta_figure},
the line for $v_{\rm hot}=50\kms$ (with diamond symbols) is mostly below the
shaded region (at $v_{\rm disk}\ga 10\kms$). The relatively low feedback
efficiency results in more efficient metal enrichment in the disks, and thus
the correlation shifts upwards along the metallicity compared with that
obtained from the fiducial model.  The slope in the feedback efficiency
($-3.2$) is steeper than the slope of the dashed line ($-2$), thus the obtained
slope at the low masses of the correlation is steeper than that obtained from
the fiducial model. Note that the linear correlation is broken at $M_*\sim 10^7
\msun$ and the slope becomes flatter above this mass. This is because
that {\it in this model} the satellites above $10^7\msun$ have
circular velocities $\ga 50\kms$ and SN feedback is ineffective with $\beta\la
1$ in these systems. Thus, the metals generated by stars are mostly left in the
galaxies, and so the total mass of metals in a galaxy is close to the the metal
yield.

Note that the satellites with stellar masses about several times $10^7\msun$
have circular velocities $\sim 30\kms$ (with $\beta \gg 1$) in the fiducial
model but which rise to $\sim 50\kms$ (with $\beta\sim 1$) in the model with
$v_{\rm hot}=50\kms$. This is because with weaker feedback, the galaxies with a
given mass tend to form in smaller halos.  The reason that a galaxy with a
given stellar mass has a higher circular velocity in a smaller halo can be
understood as follows. Considering
\begin{equation}
 j_{\rm H}=\frac{G\lambda_{\rm H} M_{\rm H}^{3/2}}{|E_{\rm H}|^{1/2}}
\end{equation}
and
\be E_{\rm H}=-\frac{GM_{\rm H}^2}{2r_{\rm vir}}, \ee
where $j_{\rm H}$ is the halo specific angular momentum, $\lambda_{\rm H}$ is
the halo spin parameter (following a log-normal distribution with mean value of
$\ln\lambda_{\rm H}$ 0.039 and its dispersion 0.53, see
\citealt{CL96,LK99,Bett07}), $M_{\rm H}$ is the mass of the dark matter halo,
$E_{\rm H}$ is the total energy of the halo, and $r_{\rm vir}$ is the virial
radius of the dark matter halo, one has
\begin{equation}
 j_{\rm d} \sim j_{\rm H} \propto \lambda_{\rm H}\sqrt{GM_{\rm H}r_{\rm vir}},
\label{eq:jdH}
\end{equation}
where $j_{\rm d}$ is the specific angular momentum of the disk, and so if a
galaxy forms in a smaller halo it has lower specific angular momentum. Further
considering
\begin{equation}
 j_{\rm d} \propto r_{\rm d}v_{\rm d} \propto \sqrt{GM_{\rm gal} r_{\rm d}},
\label{eq:jd}
\end{equation}
then one concludes that with a smaller $j_{\rm H}$, and a given galaxy mass
$M_{\rm gal}$, the size of the galaxy $r_{\rm d}$ is smaller and the circular
velocity $v_{\rm d} \sim \sqrt{GM_{\rm gal}/r_{\rm d}}$ is larger.
\end{itemize}

\begin{figure*} \center
\includegraphics[scale=1.0]{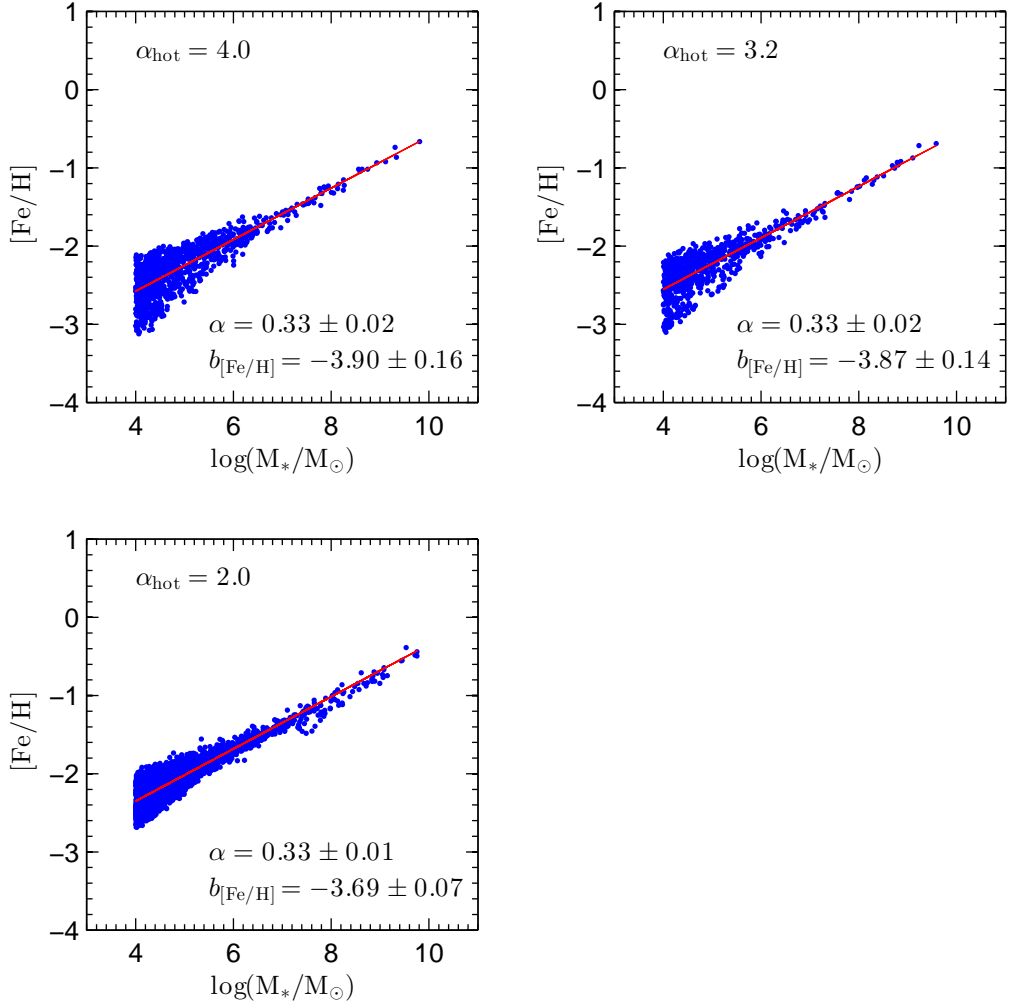}
 \caption{The stellar metallicity -- stellar mass correlations obtained from
models with different $\alpha_{\rm hot}$. All other parameters are the same as
those in the fiducial model. The line and the points have the same meaning as
those in Figure~\ref{fig:m_z_cor_vhot}.  The models with different $\alpha_{\rm
hot}$ give the almost similar results, although the model with $\alpha_{\rm
hot}=2.0$ (bottom panel) shows a slightly higher iron fraction (or a higher
best-fit intercept) and a smaller
scatter at the low-mass end. The SN feedback efficiency in these models is
mostly limited by the energy condition (see Eq.~\ref{eq:energy_condition} and
Fig.~\ref{fig:beta_figure}), although there could be some exceptions for the
model with $\alpha_{\rm hot}=2.0$.  See details in Section~\ref{sec:depvhot}.
} \label{fig:m_z_cor_ahot}
\end{figure*}

Figure \ref{fig:m_z_cor_ahot} shows the stellar mass -- metallicity
correlations of models with different values of $\alpha_{\rm hot}$.  As seen
from the figure, the models with $\alpha_{\rm hot}=4.0$ and $\alpha_{\rm
hot}=3.2$ give the same correlation, with the same slope, intercept, and
scatter. The model with $\alpha_{\rm hot}=2.0$ provides a correlation with the
same slope as the previous two, though with a little higher intercept and
smaller scatter.

The similarity in the correlations obtained with the different values of
$\alpha_{\rm hot}$ can be understood from Figure~\ref{fig:beta_figure}.  As
seen from the figure, all the lines with $\alpha_{\rm hot}=4.0,3.2,2.0$ and
$v_{\rm hot}=200\kms$ are located in the shaded region, where the feedback
predicted by the scaling law is too strong so that the effects of the feedback
is largely determined by the energy condition and produces almost the same
correlation. In Figure~\ref{fig:beta_figure}, the line with $\alpha_{\rm
hot}=2.0$ and $v_{\rm hot}=200\kms$ is close to the boundary of the shaded
region, and considering that $v_{\rm vir}=v_{\rm disk}$ is only an
approximation for drawing the dashed line, it is likely that in some cases the
feedback efficiency given by the scaling law with $\alpha_{\rm
hot}=2.0$ and $v_{\rm hot}=200\kms$ is weaker than the limit constrained by the
energy condition, thus the metal enrichment in the stars is enhanced compared
to that constrained by the energy condition. This enhancement can be seen from
the slightly higher intercept of the correlation shown in the panel with
$\alpha=2.0$ in Figure~\ref{fig:m_z_cor_ahot}.

As seen from the Figures~\ref{fig:m_z_cor_vhot} and \ref{fig:m_z_cor_ahot}, the scatter of the simulated correlation is generally
large at the low-mass end. The large scatter at the low-mass end is caused
mainly through the scatter in the star formation durations of the galaxies at a
given stellar mass and the difference in the chemical enrichment of SNe Ia and
II. If the duration is short, the metal enrichment is mainly contributed by SN
II explosion, which has a chemical pattern with a relatively low iron fraction;
while if the duration is long enough, SNe Ia may have a non-negligible
contribution to the metal enrichment and generate more iron than SNe II. Thus a
short star formation duration would lead to a low [Fe/H], while a longer one
leads to a higher [Fe/H], which causes a dispersion in the stellar
metallicities of the satellites at a given stellar mass.  Recent observations
by \citet{V13} also suggests the importance of SN Ia enrichment in some MW
ultra-faint dwarf satellites by their alpha element abundance.  Note that the
dispersion of the high-mass systems is not that large, as they all experience
extended star formation durations. The model with $v_{\rm hot}=50\kms$ in
Figure~\ref{fig:m_z_cor_vhot} shows a relatively large scatter for the
relatively high-mass systems, which is because the SN feedback is not effective
enough in that model and the scatter of the simulated present-day gas fractions
at a given stellar mass can be large.

\subsubsection{Effects of the reionization of the universe} \label{sec:depreion}

\begin{figure*}
 \includegraphics[scale=1.0]{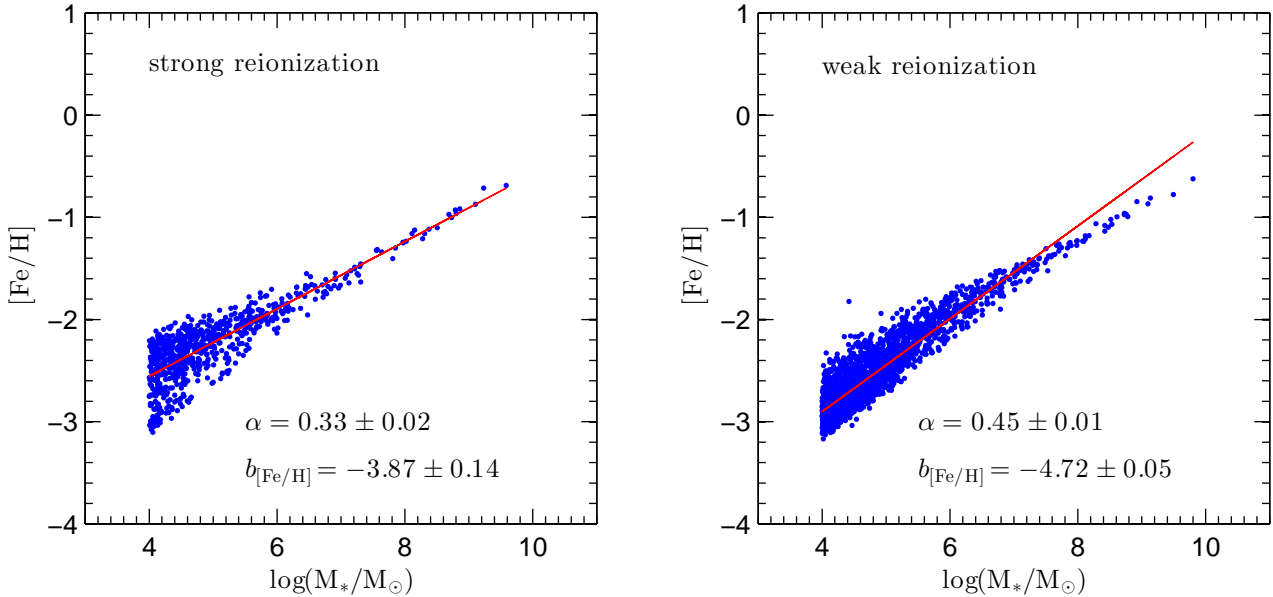}
 \caption{The stellar metallicity -- stellar mass correlations of the MW
dwarfs obtained from models with different reionization strength. The left
panel gives the result of the fiducial model; 
and the right panel shows the result for the model with all other parameters,
except for the reionization strength, being the same as those in the fiducial model. The line and the points have the same meaning as those in
Figure~\ref{fig:m_z_cor_vhot}. As seen from the figure, reducing the
reionization strength leads to an increase of the slope of the correlation.
See details in Section~\ref{sec:depreion}.}
\label{fig:m_z_reion}
\end{figure*}

\begin{figure}
\center
 \includegraphics[scale=1.0]{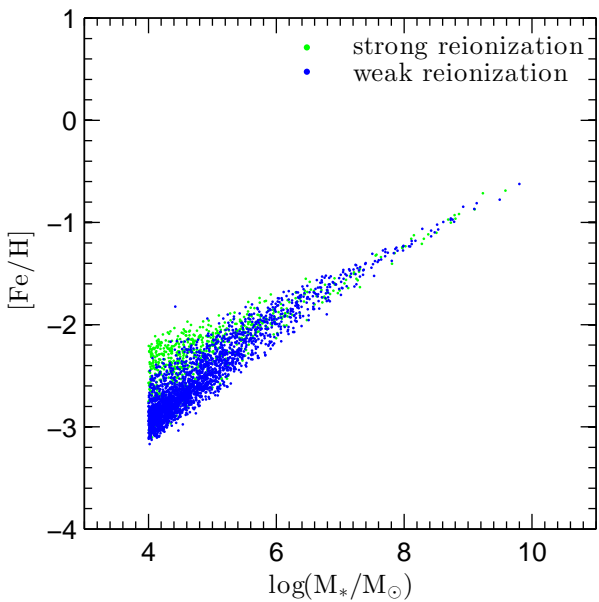}
 \caption{Comparison of the stellar mass -- metallicity correlations obtained
from the different reionization models. The blue dots are the results of the
weak reionization model, and the green dots are obtained from the strong
reionization model. As seen from the figure, the change of the reionization
strength mainly affects the metallicity of the low-mass galaxies at
$M_*\la 4\times 10^6\msun$. See details in Section~\ref{sec:depreion}.
\label{fig:m_z_cmp}}
\end{figure}

Figure~\ref{fig:m_z_reion} shows the stellar mass -- metallicity correlations
obtained in the models with strong and weak reionization. As seen from the
figure, the weak reionization model results in a steeper slope with
$\alpha\simeq 0.45$, in contrast with $\alpha\simeq 0.33$ given by the strong
reionization model.  The difference can be understood through the following
points.
\begin{itemize}
\item Compared to weak reionization, the strong reionization can heat up the
IGM with increasing pressure and also evaporate relatively more gas from a
given halo, and thus less gas in the halo cools to form a galaxy. Thus given a
galaxy mass, the halo where the galaxy forms is likely to be less massive in
the weak reionization model than that in the strong reionization model.
According to Equations (\ref{eq:jdH}) and (\ref{eq:jd}), the decrease of the
halo mass in the weak reionization model decreases the specific angular
momentum of the galaxy disk, and thus the size of the formed disk is smaller
with larger circular velocity, which would increase the star formation rate and
shorten the star formation duration. As mentioned in Section~\ref{sec:depvhot},
the shortening of the star formation duration would decrease the metallicities
of the dwarfs.
\item
For the low-mass systems, the SN feedback efficiency is determined by the limit
from the energy condition, as mentioned before. As
$v_{\rm vir}=\sqrt{GM_{\rm H}/r_{\rm vir}} \propto \sqrt{GM_{\rm H}/M_{\rm
H}^{1/3}} \propto M_{\rm H}^{1/3}$,
a smaller halo mass has a smaller virial velocity and leads to a stronger
feedback efficiency, which also decreases the metallicities of the dwarfs.
\item Note that the reionization only affects low-mass halos strongly.
Figure~\ref{fig:m_z_cmp} shows the results of the strong and the weak
reionization models together, where the correlations obtained from the two
reionization models become the same above some mass ($\sim 4\times 10^6 {\rm
M}_{\odot}$).
\end{itemize}


\subsubsection{Effects of molecular hydrogen cooling} \label{sec:depH2cooling}

\begin{figure*} \center
 \includegraphics[scale=1.0]{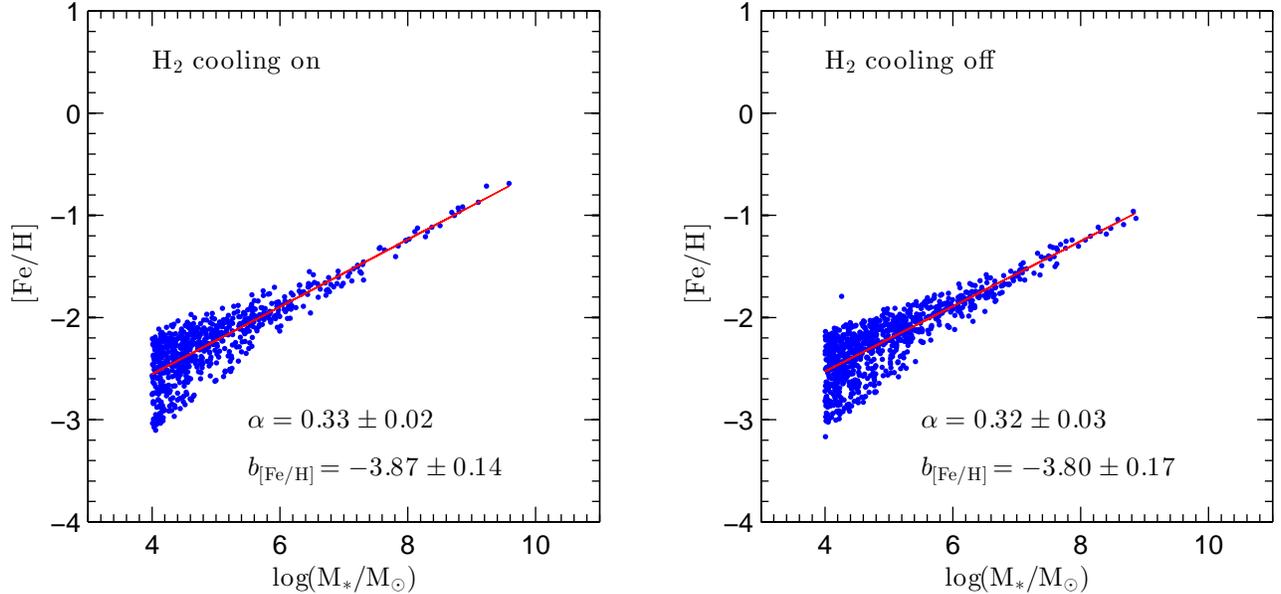}
 \caption{The stellar mass -- metallicity correlations obtained from models
with including ${\rm H}_2$ cooling (on) and excluding it (off). All the other
parameters are the same as those in the fiducial model. The line and the points
have the same meaning as those in Figure~\ref{fig:m_z_cor_vhot}. The figure
shows that the ${\rm H}_2$ cooling occurred in the early universe has little
effects in the correlation. See also Section~\ref{sec:depH2cooling}.}
\label{fig:z_m_corr_H2}
\end{figure*}

Figure \ref{fig:z_m_corr_H2} shows the correlations obtained from the models
with and without including the ${\rm H}_2$ cooling in the early universe. As
seen from the figure, these correlations are almost the same, insensitive to
the molecular hydrogen cooling process. The reason is that molecular hydrogen
cooling can only occur before the completion of the reionization when the UV
background is not too strong, and it can be important only for small halos
where the atomic cooling of gas is not effective; thus the stars formed through
the cold gas cooled by molecular hydrogen cooling are only a very small part
($<5\%$) of the final stellar populations of a given satellite, which cannot
affect the average stellar metallicity.

\begin{figure*} \center
\includegraphics[scale=0.5]{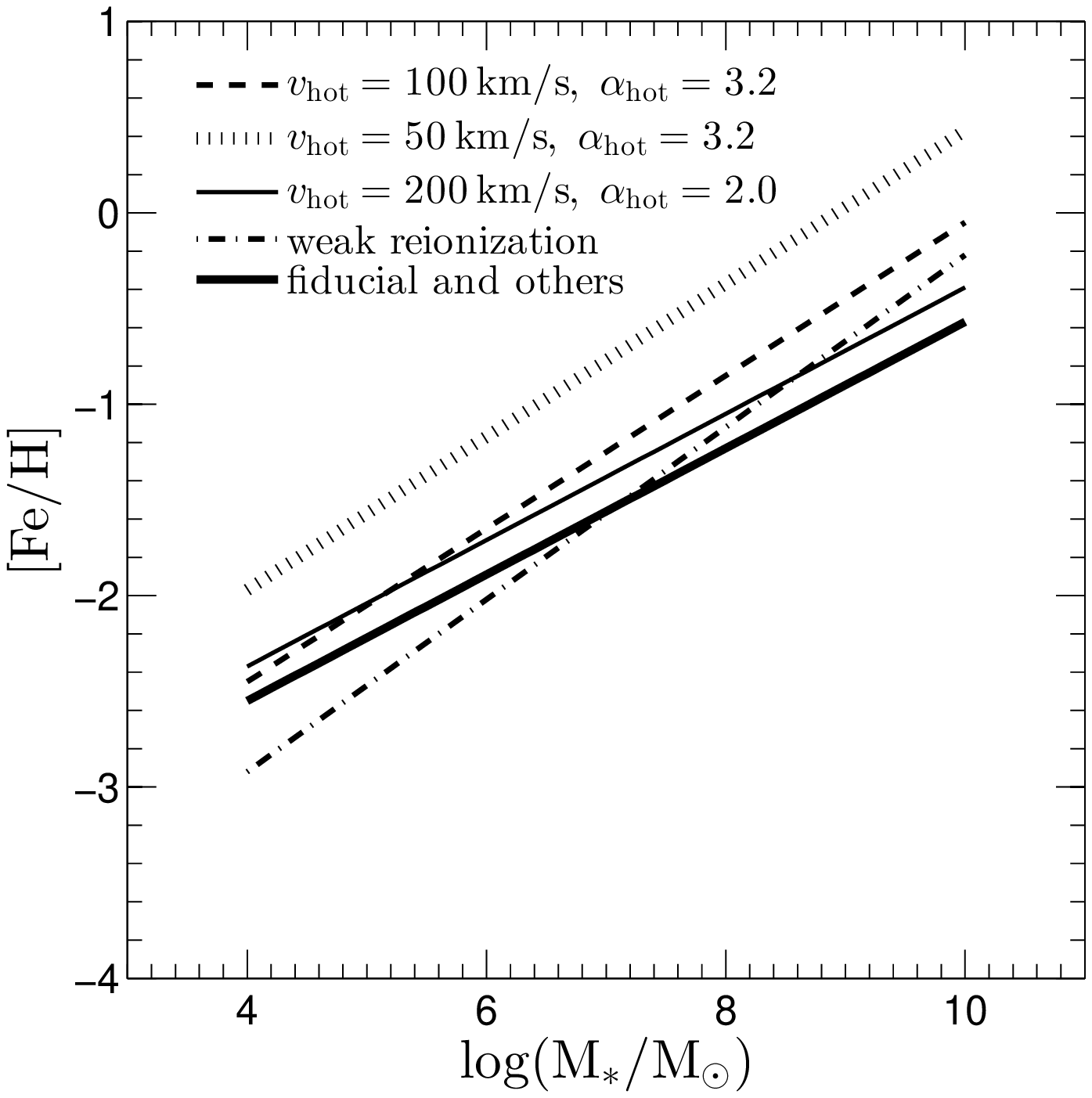}
 \caption{A summary of all the best-fit lines for the stellar mass --
metallicity correlations obtained from the models shown in
Figures~\ref{fig:m_z_cor_vhot}--\ref{fig:z_m_corr_H2}. The model parameters are
labeled in the panel.  The best-fit lines of some model parameter sets
[$(v_{\rm hot},\alpha_{\rm hot})=(400\kms,3.2)$, $(v_{\rm hot},\alpha_{\rm
hot})=(200\kms,4.0)$, and the one with excluding H$_2$ cooling] are hard to
distinguished from that of the fiducial model visually, and we use the result
of the fiducial model (thick solid line) to represent all of them.}
\label{fig:bestfits}
\end{figure*}

For view clarity, a summary of all the best-fit lines for the stellar mass
--metallicity correlations obtained from the models shown in
Figures~\ref{fig:m_z_cor_vhot}--\ref{fig:z_m_corr_H2} is given in
Figure~\ref{fig:bestfits}. 

\subsection{The metallicity distribution of individual
satellites}\label{sec:z_distribution}

\begin{figure*} \center
\includegraphics[scale=1.0]{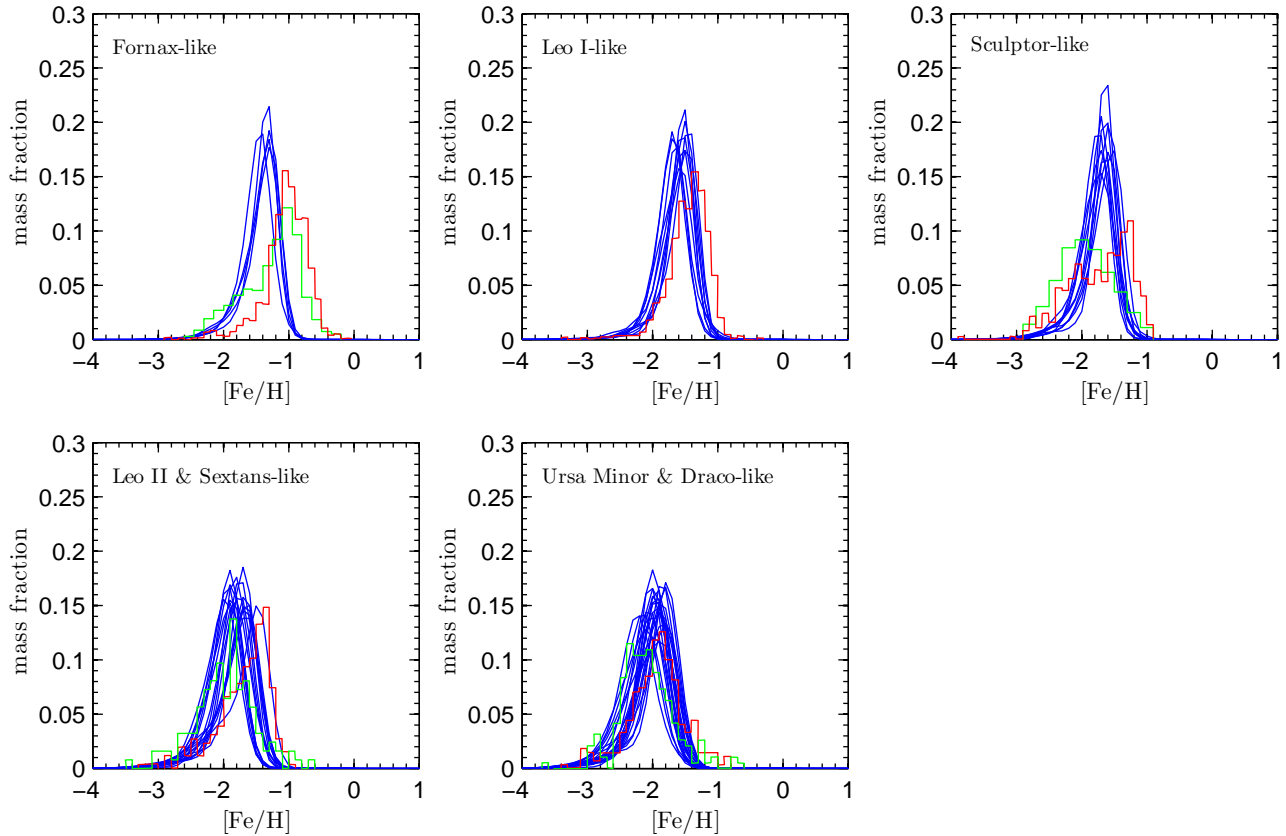}
\caption{The metallicity distributions of individual satellites. Different
panels represent the satellites with different luminosity ranges. The blue
curves represent our simulation results. The red histograms in all the panels
and the green histograms in the bottom panels represent the observational
distributions obtained from \citet{kirby1}; and the green histograms in the top
`Fornax-like' and `Sculptor-like' panels represent the observational
distributions obtained from \citet{fornax} and \citet{sculptor}, respectively.
The simulated distributions shown in the figure are obtained by convolving with
a Gaussian function to model the observational errors, where the medians of the
errors of the seven classical dSphs shown in \citet{kirby1} is adopted, i.e.,
$0.12$ dex for Fornax-like, Leo I-like, and Sculptor-like satellites, $0.15$
dex for Leo II and Sextans-like satellites, and $0.18$ dex for Ursa Minor and
Draco-like satellites.  Our simulations roughly reproduce the observational
results, except for the distributions in Fornax and Sculptor.  See
Section~\ref{sec:z_distribution} for more discussion.  \label{fig:z_d}}
\end{figure*}

Figure~\ref{fig:z_d} shows the metallicity distributions of the satellites
generated by the fiducial model. The simulated satellites are classified
according to their luminosities. The simulation results are shown in blue
curves, and the red histograms in all the panels and the green histograms in
the bottom panels represent the observational results in \citet{kirby1}. The
five panels represent different $V$-band magnitude bins.  From left top to
right bottom, the five bins are classified as Fornax-like ($-13.5 < M_V \le
-12.5$), Leo I-like ($-12.5 < M_V \le -11.5$), Sculptor-like ($-11.5 <M_V \le
-10.5$), Leo II and Sextans-like ($-10.0 < M_V \le -9.2$) and Ursa Minor and
Draco-like ($-9.2 < M_V \le -8.5$), respectively.  In general, our simulations
reproduce the observations, except for the high-metallicity peak of Fornax and
the double peaks of Sculptor.  However, there are also some observations
supporting that Fornax has a lower metallicity, e.g., the peak being at
[Fe/H]$\sim-1.3$ in the photometric metallicity distribution of
\citet{stetson1998}, which agrees with the simulation results
better.  Recent observation by \citet{fornax} also gives a lower average
metallicity for Fornax (green histogram in the top `Fornax-like' panel), but it
is still higher than our simulation result.  For Sculptor, currently it is the
only satellite showing a double-peaked metallicity distribution, so it is
possible that this distribution comes from a special formation history which is
too rare to be realized in the about $10$ MW-like hosts. In addition, the
observational metallicity distribution of Sculptor reported in \citet{sculptor}
has only one peak (see also \citealt{helmi2006}), as shown by the green diagram
in the top `Scupltor-like' panel, which leads to a better agreement with our
simulation results; even if the observational double-peaked distribution is
true, the simulation result still captures some basic observational chemical
characteristics of Sculptor, as they are in the same metallicity range in
Figure~\ref{fig:z_d}.

We introduce the following three quantities to characterize the metallicity
distributions quantitatively.
\begin{itemize}
\item The first one is the corresponding [Fe/H] value at the peak of a
metallicity distribution.
\item The second one is based on the so-called linear metallicity dispersion,
$\sigma (\calZ)$, which is defined by (cf., \citealt{Leaman2012})
\begin{equation}
\sigma (\calZ)\equiv \left[\int (\calZ-\bar{\calZ})^2 P(\calZ) d\calZ \right]^{1/2},
\label{eq:sigma_z}
\end{equation}
where $\calZ\equiv10^{\rm [Fe/H]}$, $P(\calZ)$ is the metallicity distribution
and $P(\calZ)d\calZ$ represents the mass fraction of the stars with metallicity
in the range $\calZ\rightarrow \calZ+d\calZ$, and
\begin{equation}
\bar{\calZ}\equiv\int \calZ P(\calZ) d\calZ. \label{eq:z_bar}
\end{equation}
The quantity of $\sigma(\calZ)$ mainly evaluates the extension of the
metal-rich tail of a metallicity distribution, because $\calZ$ gives a
relatively large weight to metal-rich stars for calculating $\sigma(\calZ)$.
As $\sigma (\calZ)$ would be larger with larger $\bar{\calZ}$, we employ the
relative dispersion, $\sigma (\calZ)/\bar{\calZ}$, to represent the intrinsic
extension of the metal-rich tails.
\item The third one is also based on a dispersion, $\sigma(\calZ')$, where
$\calZ'\equiv\sqrt{1/10^{\rm [Fe/H]}}$, defined by the following equations
similar to Equations (\ref{eq:sigma_z}) and (\ref{eq:z_bar}):
\begin{equation}
\sigma (\calZ')\equiv \left[\int (\calZ'-\bar{\calZ'})^2 P(\calZ')
d\calZ'\right]^{1/2}
\end{equation}
and
\begin{equation}
\bar{\calZ'}\equiv \int \calZ' P(\calZ') d\calZ',
\label{eq:barZp}
\end{equation}
where $P(\calZ')$ is the distribution function of $\calZ'$ and
$P(\calZ')d\calZ'$ represents the mass fraction of the stars with $\calZ'$ in
the range $\calZ'\rightarrow \calZ'+d\calZ'$.  The quantity of $\calZ'$ gives a
large weight to metal-poor stars and hence $\sigma(\calZ')$ mainly represents
the extension of the metal-poor tail of a metallicity distribution. Because
$\sigma (\calZ')$ would be larger with larger $\bar{\calZ'}$, we employ the
relative dispersion, $\sigma (\calZ')/\bar{\calZ'}$, to represent the intrinsic
extension of the metal-poor tails. Note that for an appropriate representation
of the metal-poor tails, we are careful in choosing the definition of $\calZ'$,
for example, a square root is introduced into the definition to avoid the
domination of the extremely metal-poor stars with [Fe/H]$<-5$ in the value of
$\calZ'$, as the statistics of the extremely metal-poor stars is uncertain due
to possible non-homogeneous mixing and turbulent dynamics of gas during their
formation processes.
\end{itemize}
The corresponding [Fe/H] value at the peak of a metallicity distribution is
roughly the same as the average [Fe/H] value, and the behavior of the average
has been analyzed in Section~\ref{sec:sub1}, so in the following analysis we
only focus on the two relative dispersions, $\sigma (\calZ)/\bar{\calZ}$ and
$\sigma (\calZ')/\bar{\calZ'}$, in Sections~\ref{sec:metalpoor} and
\ref{sec:metalrich}, respectively.

Figure \ref{fig:example} shows an example of $\sigma (\calZ)/\bar{\calZ}$ and
$\sigma (\calZ')/\bar{\calZ'}$ obtained from the fiducial model, in which both
the original results from the simulations (small colored dots) and the
statistical results (black solid circles and their error bars) are presented.
As seen from the figure, the statistical results represent the general trends
of the original results obtained from the simulations well. For simplicity and
clarity, in the following similar figures indicating $\sigma
(\calZ)/\bar{\calZ}$ and $\sigma (\calZ')/\bar{\calZ'}$
(Figs.~\ref{fig:z_d_vhot}--\ref{fig:z_d_h2}), we only show the statistical
results of the solid circles and their error bars, but omit the original
results of the small dots. 

\subsubsection{Metal-poor tails} \label{sec:metalpoor}

We find that the metal-poor tails of the satellites in most of their stellar
mass ranges are mainly constructed through minor mergers, except for some small
satellites (with $M_*\sim 10^3$--$10^5\msun$). In general, the metallicity
distribution of any galaxy has a metal-poor tail, and the stars in the
metal-poor tail may come from star formation in the galaxy itself or from an
accreted galaxy. We define a dispersion ratio of the metal-poor tails and show
the logarithm of the ratio in the right panel of Figure 11 by color scales to
indicate the relation of the metal-poor tail of a satellite with the stars
accreted onto it in previous minor mergers occurred before the progenitor of
the satellite falls into a big halo to become the satellite. Here the
dispersion ratio of the metal-poor tails is defined by the ratio of
$\sigma_{\rm remov}(\calZ')/\sigma(\calZ')$, where $\sigma(\calZ')$ is derived
from the stellar metallicity distribution of a present-day satellite, and
$\sigma_{\rm remov}(\calZ')$ is the dispersion of the metal-poor tail derived
from the stellar metallicity distribution which is obtained by removing the
stars accreted onto the satellite in previous minor mergers.  As seen from the
color distribution of the points in the right panel of the figure, the
satellites at most of their stellar mass ranges have a low dispersion ratio,
which indicates that minor mergers have a significant contribution to their
metal poor tails; and only some small satellites (with $M_*\sim
10^3$--$10^5\msun$) have a high ratio, which indicates that the contribution
from minor mergers is negligible.  For those with relatively low dispersion
ratios, in minor mergers, the accreted small galaxy usually has a lower
metallicity, and the major part of its stellar population would appear in the
metal-poor tail of the merged galaxy. In the minor mergers with relatively
large mass ratios (e.g., not significantly lower than the mass ratio criterion
$1/3$ for minor/major mergers), the mass of the stars in the accreted small
galaxy is usually large than that of the metal-poor stars in the accreting
large galaxy, and thus the stars in the accreted galaxy would dominate the
properties of the metal-poor stars of the merged galaxy and hence the quantity
of $\sigma(\calZ')/\bar{\calZ'}$. 

\begin{figure*} \center
\includegraphics[scale=1.0]{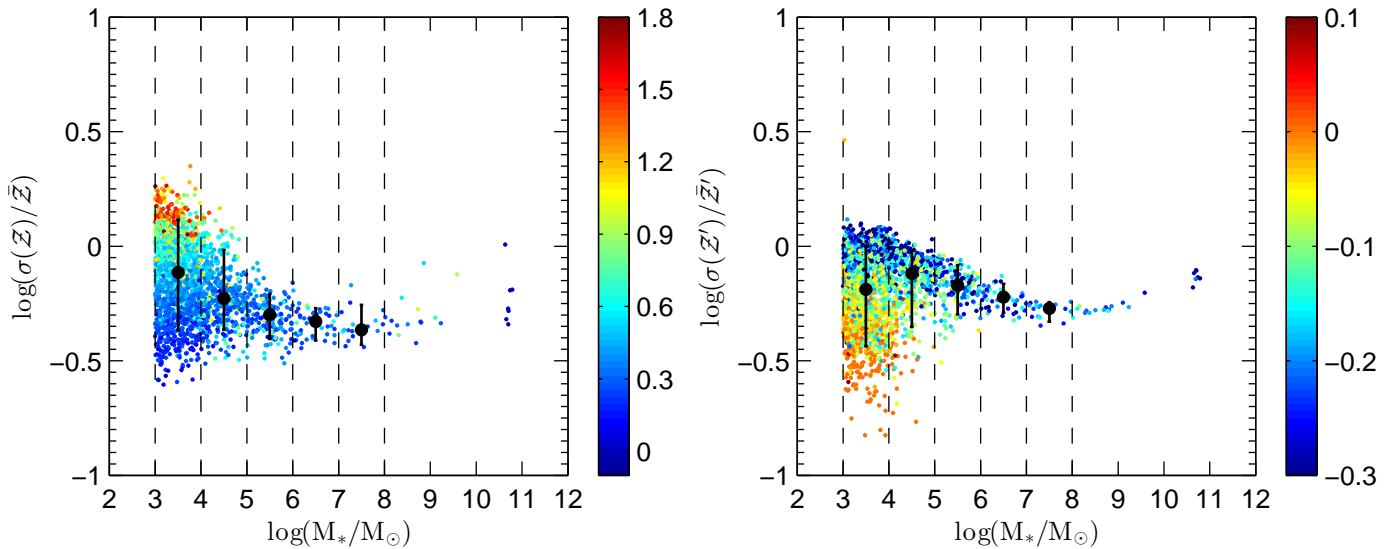}
\caption{An example for the relative metallicity dispersions $\sigma
(\calZ)/\bar{\calZ}$ and $\sigma (\calZ')/\bar{\calZ'}$ (see definitions in
Eqs.~\ref{eq:sigma_z}--\ref{eq:barZp}) obtained from the fiducial model.  The
small colored dots show the original results from the simulations. The galaxies
are further bound into five stellar mass bins, $10^3 \-- 10^4\msun$, $10^4 \--
10^5\msun$, $10^5 \-- 10^6\msun$, $10^6 \-- 10^7\msun$, and $10^7 \--
10^8\msun$, indicated by the black dashed vertical lines. In each of these mass
bins, the black solid circle represents the median of the small dots, and the
error bar of each circle represents the range bounded by the lowest 10\% and
the highest 10\% of the small dots. We do not show the statistical results for
satellites with mass higher than $10^8\msun$ due to their small numbers. As
seen from the figure, the statistical results can well present the major
features of the original results.  The color scale in the left panel represents
$\Delta$[Fe/H]$\equiv$[Fe/H]$_{\max}-\overline{\rm [Fe/H]}$, where
[Fe/H]$_{\max}$ is the maximum value of the stellar [Fe/H] when the progenitor
of a satellite falls into a big halo and $\overline{\rm [Fe/H]}$ is the average
[Fe/H] at that time.  The color scale in the right panel represents 
$\log[\sigma_{\rm remov}(\calZ')/\sigma(\calZ')]$, where $\sigma_{\rm
remov}(\calZ')$ is the dispersion of the metal-poor tail derived from the
stellar metallicity distribution which is obtained by removing the stars
accreted onto the satellite in previous minor mergers.  The color distributions
in the left panel indicates that the relative extension of the metal-rich tail
of a satellite is usually formed before the progenitor of the satellite falls
into a big halo to become the satellite, and the star formation after infall
has negligible effects to the final shape of the metal-rich tail.  The color
distributions in the right panel indicates that the metal-poor tails of the
satellites in most of their stellar mass ranges are mainly constructed through
minor mergers, except for some small satellites (with $M_*\sim
10^3$--$10^5\msun$). See details in Section~\ref{sec:z_distribution}.
} \label{fig:example}
\end{figure*}

Some tendencies of $\sigma(\calZ')/\bar{\calZ'}$ with different satellite
masses are illustrated through the example of $\sigma(\calZ')/\bar{\calZ'}$
shown in the right panel of Figure~\ref{fig:example}.
\begin{itemize}
\item The maximum $\sigma(\calZ')/\bar{\calZ'}$ declines with increasing galaxy
stellar mass.  This can be understood as follows. Almost all the large
$\sigma(\calZ')/\bar{\calZ'}$, which form the upper boundary in
Figure~\ref{fig:example}, are caused by minor mergers.  The galaxies with low
stellar masses usually formed earlier; and at the earlier time, minor mergers
are more likely to have larger mass ratios, e.g.\ $1/5$, and thus the accreted
galaxies have a larger contribution to the metal-poor tails of the merged
galaxy.
\item For relatively large satellites (e.g., with stellar mass above
$10^{5.5}\msun$), the scatter in the values of $\sigma(\calZ')/\bar{\calZ'}$ is
small.  For large satellites, their metal-poor tails are all constructed by
minor mergers.  Due to their relatively large stellar mass, it usually took a
long time for them to form, and they would have accreted many small and
metal-poor galaxies during their formation histories; and thus the properties
of their metal-poor tails are determined by the sum of the metallicities of
those accreted small galaxies.  Although the metallicity of each individual
accreted galaxies could be quite different, their sum has a relatively smaller
scatter according to the central limit theorem.  Furthermore, the average
metallicities of these relatively large galaxies have a very small scatter as
shown in the stellar metallicity versus stellar mass correlations above. Hence
the scatter of the $\sigma(\calZ')/\bar{\calZ'}$ of these relatively large
galaxies is small.
\item For relatively small satellites (e.g., with stellar mass below
$10^{5.5}\msun$), the scatter in the values of $\sigma(\calZ')/\bar{\calZ'}$ is
large.  For the small satellites, their formation durations are usually short,
and not all of them have enough time to have experienced a minor merger. If a
galaxy did not experience a minor merger, then the metal-poor tail is
constructed by the stars formed in itself.  Thus the extension of the
metal-poor tail, or equivalently the value of the
$\sigma(\calZ')/\bar{\calZ'}$, is correlated to the star formation rate of this
galaxy.  With a higher star formation rate, the $\sigma(\calZ')/\bar{\calZ'}$
would be lower, as less metal can involve into star formation and the stars
tend to have the same metallicity. For these small galaxies, they can be formed
with both large and small star formation rate, so the diversity of the
$\sigma(\calZ')/\bar{\calZ'}$ value is large.
\end{itemize}

The dependence of the $\sigma(\calZ')/\bar{\calZ'}$ on the parameters of
different physical processes is illustrated in the right panels of
Figures~\ref{fig:z_d_vhot}--\ref{fig:z_d_h2}. As seen from the figures, the
statistical results of $\sigma(\calZ')/\bar{\calZ'}$ are not sensitive to the
used feedback parameters and the molecular hydrogen cooling process.

Figure~\ref{fig:z_d_reion} shows the statistical results of $\sigma
(\calZ')/\bar{\calZ'}$ for different reionization models. As seen from the
right panel, $\sigma (\calZ')/\bar{\calZ'}$ in the low-mass range $10^3 \--
10^5\msun$ is lower in the weak reionization model than that in the strong
reionization model, but $\sigma (\calZ')/\bar{\calZ'}$ in the mass range $10^6
\-- 10^8\msun$ is higher in the weak reionization model than that in the strong
reionization model. The reason for a lower $\sigma (\calZ')/\bar{\calZ'}$ in
the low-mass range $10^3 \-- 10^5\msun$ resulting from the weak reionization
model is that the host halos of those galaxies are smaller and thus experience
fewer mergers than the corresponding ones in the strong reionization model.
Furthermore, smaller halos lead to smaller disk specific angular momenta and
higher star formation rates as mentioned in Section~\ref{sec:sub1}.  Therefore,
the metal-poor tail of a galaxy is weakened in the weak reionization model
compared with that in the strong reionization model, no matter whether this
tail is formed through minor mergers or star formation in itself.  For galaxies
in the mass range $10^6 \-- 10^8\msun$, their metal-poor tails have significant
contributions from the accreted smaller galaxies, as mentioned above. Although
the enhancement of the SN feedback strength and the star formation rate in the
small galaxies ($10^3 \-- 10^5\msun$) reduces their stellar metallicities, the
different reionization models cause limited effects on the average
metallicities in larger galaxies ($10^6 \-- 10^8 \msun$), so the difference in
the metallicities of the larger galaxies and the smaller accreted galaxies is
enhanced by adopting weaker reionization, which prolongs the metal-poor tails
of the larger galaxies.

\subsubsection{Metal-rich tails} \label{sec:metalrich}

We find that in the fiducial model, the relative extension of the metal-rich
tail of a satellite is usually formed before the progenitor of the satellite
falls into a big halo to become the satellite, and the star formation after
infall has negligible effects to the final shape of the metal-rich tail.
We define $\Delta$[Fe/H]$\equiv$[Fe/H]$_{\max}-\overline{\rm [Fe/H]}$ and show
it in the left panel of Figure~\ref{fig:example} by color scales to indicate
the relation of the metal-rich tails of the satellites with infall, where
[Fe/H]$_{\max}$ is the maximum value of the stellar [Fe/H] when the progenitor
of the satellite falls into a big halo and $\overline{\rm [Fe/H]}$ is the
average [Fe/H] at that time (note that the definition of such a variable to
illustrate this relation is not unique).  As seen from the color distribution
of the points in Figure~\ref{fig:example}, most of the satellites with low
$\sigma({\calZ})/\bar{\calZ}$ have a low $\Delta$[Fe/H] and those with high
$\sigma({\calZ})/\bar{\calZ}$ have a high $\Delta$[Fe/H].  Only some
small satellites (with $M_*\sim 10^3$--$10^5\msun$) have a high $\Delta$[Fe/H].
For those with low $\Delta$[Fe/H] at infall, a strong-enough star formation and
chemical enrichment after infall should be required to increase the extension
of the metal-rich tails. However, after infall, the original halo of the galaxy
may be tidally disrupted in the big halo, and the effect of SN feedback is very
strong in the fiducial model.  The strong feedback would strongly restrict the
number of the stars that can be formed in the post-infall stage, which limits
the possible value and the scatter range of $\sigma(\calZ)/\bar{\calZ}$. This
is why the $\sigma(\calZ)/\bar{\calZ}$ values shown in Figure~\ref{fig:example}
appear flat over stellar masses and has small scatters for relatively large
galaxies (with stellar mass $10^5 \-- 10^8\msun$).  The scatter in the galaxies
with stellar mass above $10^8\msun$ begins to increase, as the feedback in
these large galaxies is not as strong as that in those smaller galaxies. 

The scatter of $\sigma(\calZ)/\bar{\calZ}$ becomes very large in the very small
stellar mass range, i.e., $10^3\-- 10^5\msun$. The reason can be understood as
follows.  These small galaxies are formed in small halos and the amount of gas
for star formation is small. It is possible that the majority of the gas in
some galaxies (with high $\Delta$[Fe/H]) has turned into stars before infall,
and thus the relatively large extension of their metal-rich tails are actually
formed before infall, when SN feedback is much weaker than that after infall.
The weak feedback strength leads to a strong metal-rich tail. For the galaxies
whose metal-rich tails are not quite extended before infall, their
$\sigma(\calZ)/\bar{\calZ}$ are similar to those of larger galaxies. For larger
galaxies, their halos are larger and contains more gas, and generally they do
not turn a large part of the gas into stars before infall, so their
$\sigma(\calZ)/\bar{\calZ}$ are not large.

\begin{figure*}
\center
\includegraphics[scale=1.0]{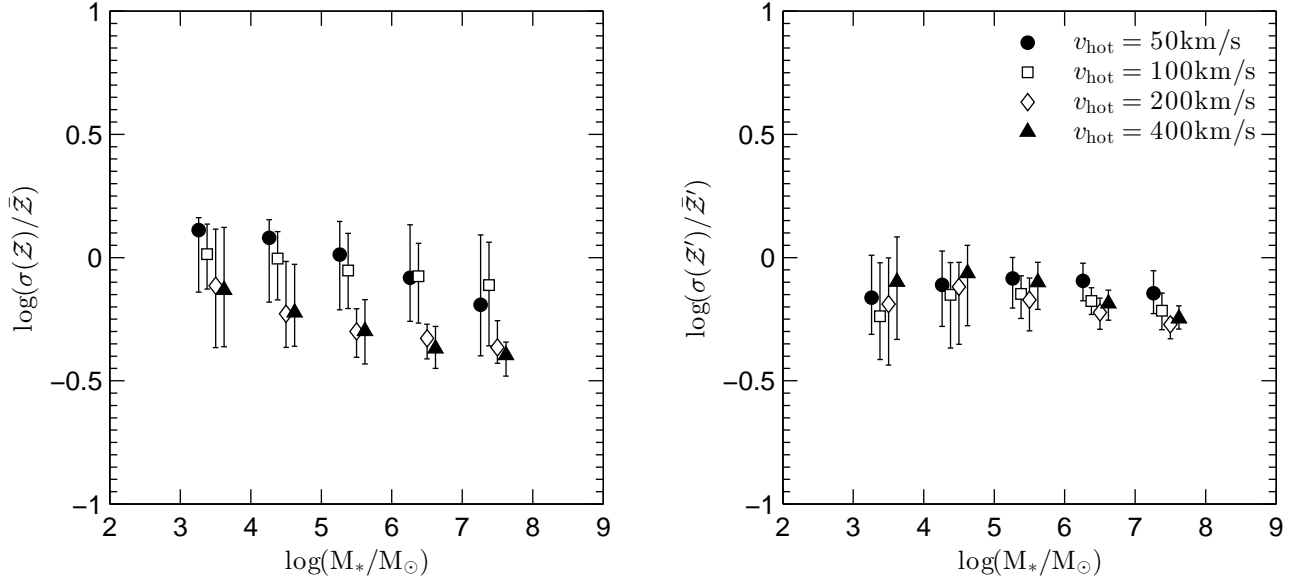}
\caption{Metallicity dispersions for different $v_{\rm hot}$. The parameters
other than $v_{\rm hot}$ all have the same values as those for the fiducial
model. The points and the error bars have the same meanings as those in
Figure~\ref{fig:example}. For view clarity, the points in the same stellar mass
bin are shifted horizontally a little one another. As seen from the figure,
the metal-poor tails are not sensitive to $v_{\rm hot}$, while the metal-rich
tails are sensitive to the reduction of the SN feedback strength through
reducing $v_{\rm hot}$.  For more details, see the
caption of Figure~\ref{fig:example} and
Section~\ref{sec:z_distribution}.\label{fig:z_d_vhot}}
\end{figure*}

\begin{figure*}
\includegraphics[scale=1.0]{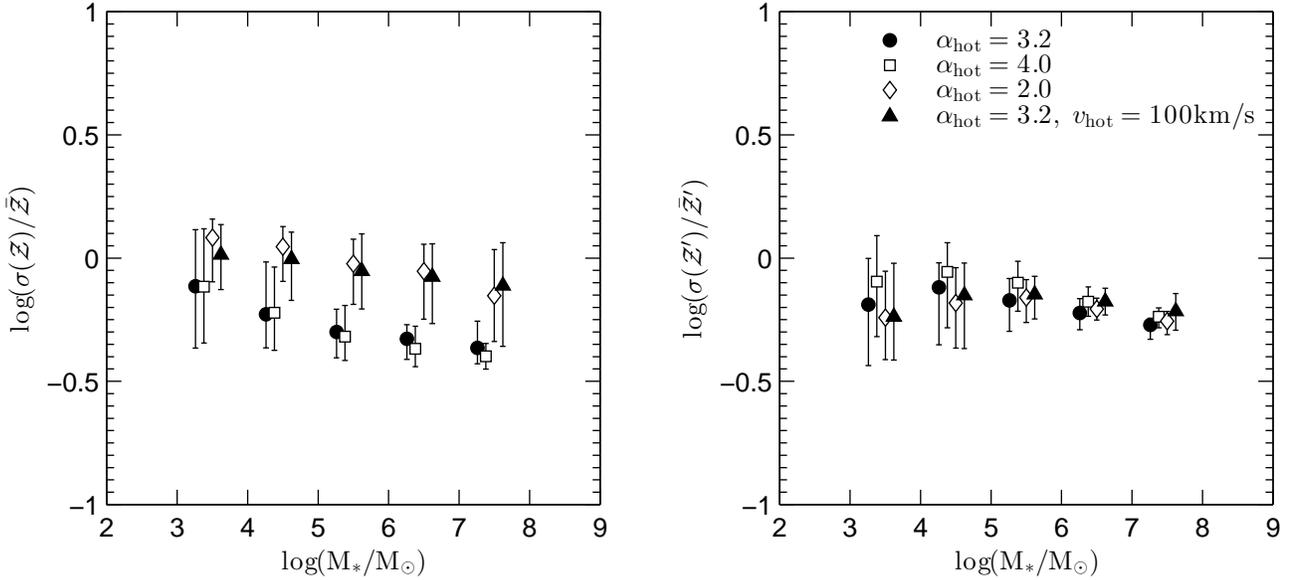}
\caption{Metallicity dispersions for different $\alpha_{\rm hot}$.  All the
other parameters not labeled in the figure have the same values as those for
the fiducial model.  The points and the error bars have the same meanings as
those in Figure~\ref{fig:example}. As seen from the figure, the metal-poor
tails are not sensitive to $\alpha_{\rm hot}$; while the metal-rich tails are
sensitive to the reduction of the SN feedback strength through reducing
$\alpha_{\rm hot}$ (note that some degeneracy exists in the results obtained
from different $v_{\rm hot}$ and $\alpha_{\rm hot}$).
For more details, see the caption of
Figure~\ref{fig:example} and Section~\ref{sec:z_distribution}.
\label{fig:z_d_ahot}}
\end{figure*}

\begin{figure*}
\center
\includegraphics[scale=1.0]{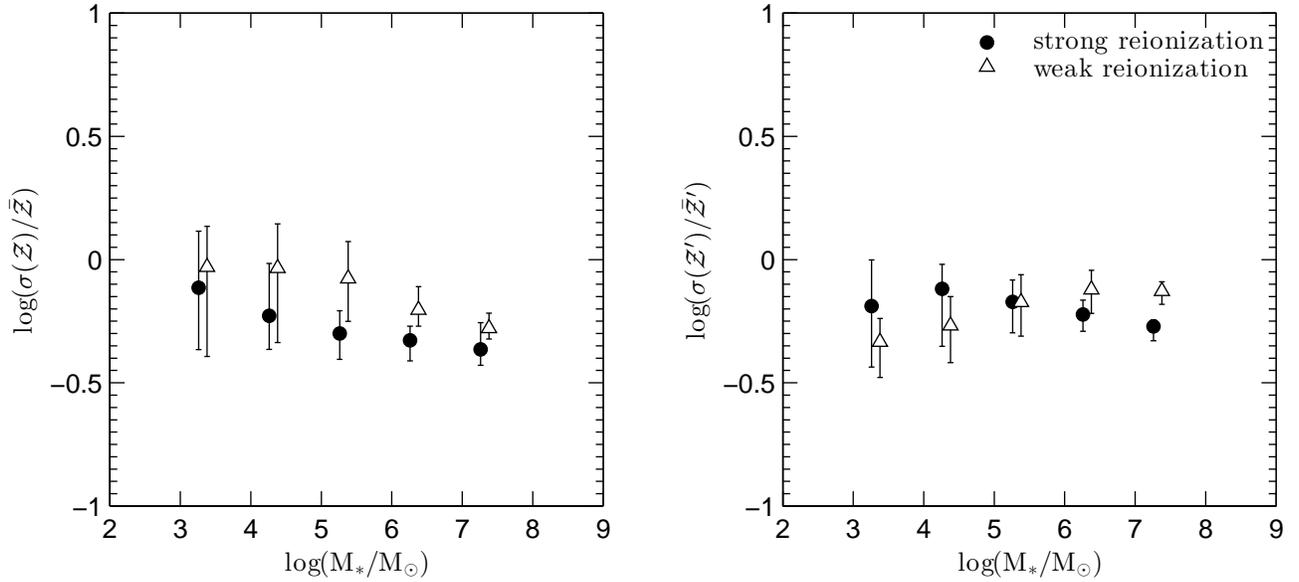}
\caption{Metallicity dispersions of different reionization models. All the
other parameters are the same as those for the fiducial model.  The points and
the error bars have the same meanings as those in Figure~\ref{fig:example}.
This figure indicates that different reionization strength can cause different
effects on both metal-poor and metal-rich tails.
For more details, see the caption of Figure~\ref{fig:example} and
Section~\ref{sec:z_distribution}.
\label{fig:z_d_reion}}
\end{figure*}

\begin{figure*} \center \includegraphics[scale=1.0]{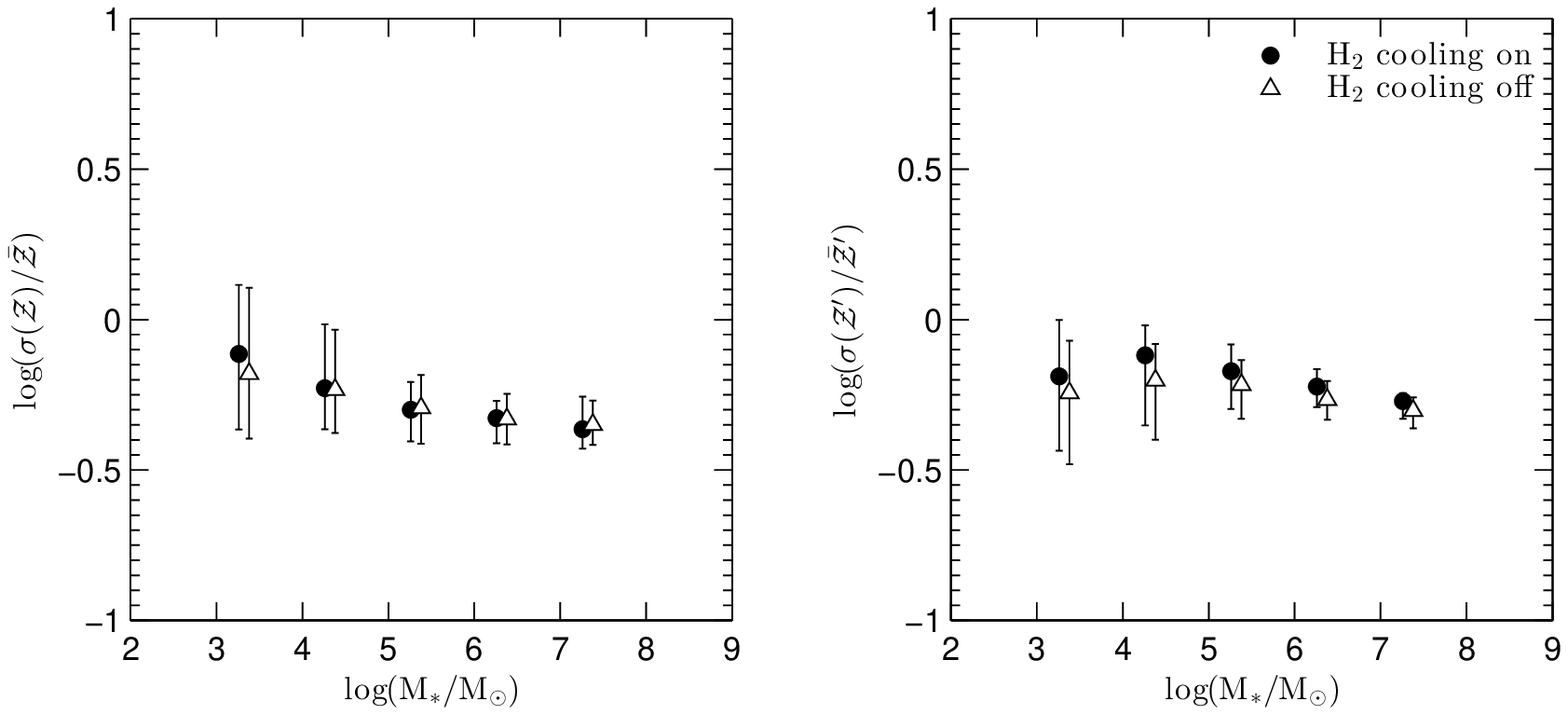}
\caption{Metallicity dispersions obtained with and without including ${\rm
H}_2$ cooling.  All the other parameters are the same as those for the fiducial
model.  The points and the error bars have the same meanings as those in
Figure~\ref{fig:example}. The figure indicates that neither metal-poor tails
nor metal-rich tails are sensitive to the ${\rm H}_2$ cooling process. For more
details, see the caption of Figure~\ref{fig:example} and
Section~\ref{sec:z_distribution}.  \label{fig:z_d_h2}}
\end{figure*}

The dependence of the metal-rich tails on different physical processes are
illustrated in the left panels of Figures~\ref{fig:z_d_vhot}-\ref{fig:z_d_h2}.
\begin{itemize}
\item Figure~\ref{fig:z_d_vhot} shows the statistical result of $\sigma
(\calZ)/\bar{\calZ}$ obtained from the models with different $v_{\rm hot}$.
The models with $v_{\rm hot}=400\kms$ and $v_{\rm hot}=200\kms$ have almost the
same dispersions, because in these two cases the feedback strength provided by
the scaling law is too strong to have significant star formation after infall.
As $v_{\rm hot}$ decreases to $100\kms$ or even $50\kms$, the effects of
different $v_{\rm hot}$ begin to appear, and the galaxies tend to have higher
$\sigma (\calZ)/\bar{\calZ}$, because the weaker feedback allows more
stars to be formed and more efficient metal enrichment. 
 
\item Figure~\ref{fig:z_d_ahot} shows the statistical result of $\sigma
(\calZ)/\bar{\calZ}$ obtained from the models with different $\alpha_{\rm
hot}$. For $\alpha_{\rm hot}=4.0$ and $\alpha_{\rm hot}=3.2$, the dispersions
are the same. This is again because the feedback in these two cases strongly
suppresses the star formation in the post-infall stage. With $\alpha_{\rm
hot}=2.0$, the feedback is weakened, which allows the formation of more
metal-rich stars, enhances the metal-rich tails of the distributions, and thus
increases $\sigma (\calZ)/\bar{\calZ}$. This figure also shows the results
obtained from the model with $\alpha_{\rm hot}=3.2$ and $v_{\rm hot}=100\kms$
for comparison, which  are similar to those of the model
with $\alpha_{\rm hot}=2.0$ and $v_{\rm hot}=200\kms$. The degeneracy of the
models with $\alpha_{\rm hot}=3.2$, $v_{\rm hot}=100\kms$ and $\alpha_{\rm
hot}=2.0$, $v_{\rm hot}=200\kms$ can be understood from
Figure~\ref{fig:beta_figure}, in which the lines representing the SN feedback
efficiency of the two models (with circles and with triangles) are close.
 
\item Figure~\ref{fig:z_d_reion} shows the statistical results of $\sigma
(\calZ)/\bar{\calZ}$ obtained from different reionization models.  As seen from
the left panel, $\sigma (\calZ)/\bar{\calZ}$ obtained in the weak reionization
model is higher than that in the strong reionization model. This can be
understood as follows. As pointed out in Section~\ref{sec:sub1}, given the
stellar mass of a galaxy, reducing reionization strength shifts its formation
into lower mass halos, and increases its disk circular velocity and the galaxy
star formation rate, which would enhance the possibility for a galaxy turning
the majority of the gas into stars before its infall, and thus enhance the
metal-rich tail. Furthermore, a larger disk circular velocity would also reduce
the feedback coefficient $\beta$ and help to enhance the star formation after
infall. For the very small galaxies ($10^3 \-- 10^4\msun$), the enhancement is
limited, because even with strong reionization, a part of them can still turn
the majority of the gas into stars before infall. But for galaxies in $10^4 \--
10^6\msun$, the enhancement of metal-rich tails is quite obvious.  This
enhancement becomes small again in more massive mass range $10^6 \--
10^8\msun$, because reionization cannot strongly affect the more massive dark
matter halos as mentioned before.
 
\item Figure~\ref{fig:z_d_h2} shows the statistical results of $\sigma
(\calZ)/\bar{\calZ}$ obtained from models with and without including molecular
hydrogen cooling processes.  These two models give almost the same dispersions,
which means that the meal-rich tails are not sensitive to the molecular
hydrogen cooling process. This is easily understood as follows. Molecular
hydrogen cooling is only expected to work before the accomplishment of
reionization, after which the UV background would strongly dissociate hydrogen
molecules and suppress this cooling mechanism. It is only important for
very small halos, and hence is unlikely to contribute to the metal-rich tails
of satellites which are formed at lower redshift and in relatively large halos.
\end{itemize}

\section{Discussions} \label{sec:discussion}

We have studied the behaviors of various metallicity properties under
different physical conditions, and these metallicity properties can be used to
put constraints on the underlying physical processes of galaxy formation,
such as reionization and SN feedback.

\subsection{Constraints on the reionization model}

\citet{Kirby13} measure the slopes of the stellar metallicity -- luminosity
correlation and the stellar metallicity -- stellar mass correlation for the MW
satellites and find $\alpha_{\rm obs} = 0.29\pm 0.02$ and $0.30\pm 0.02$,
respectively. Among all the models mentioned in Section~\ref{sec:result}, the
fiducial model with strong reionization provides a similar slope (note that the
best-fit intercept obtained from the fiducial model is also roughly consistent
with the observational correlation provided in \citealt{Kirby13}).  According
to the change tendency of the correlation slope under different physical
conditions obtained above, below we argue that it is unlikely to produce such a
slope under the weak reionization model even by varying the other parameters in
the different physical processes involved.
\begin{itemize}
\item Under the weak reionization model, if all the other parameters are the
same as those in the fiducial model, we find that the slope of the correlation
predicted by the simulation is about $0.45$, which is substantially steeper than
the observational slope.

\item The relatively steep slope predicted by the weak reionization model cannot
be reduced by adjusting the molecular hydrogen cooling process. As indicated in
Figure~\ref{fig:m_z_cor_vhot}, the ${\rm H}_2$ cooling has very limited effects
on the average stellar metallicities of larger satellites because the major
parts of their stars are formed in atomic cooling halos. For small satellites,
it enables the star formation in smaller halos, and thus leads to an earlier
enrichment. So turning off molecular hydrogen cooling can only delay the
enrichment and thus reduce the average metallicities of small galaxies. And
this would only increase the slope rather than reduce it.

\item The relatively steep slope predicted by the weak reionization model cannot
be reduced by increasing the SN feedback efficiency. As shown in
Figure~\ref{fig:beta_figure}, the feedback strength predicted by the scaling law
with the fiducial model parameters has already been very high to be limited by
the energy condition, so there is no much difference in the results by further
increasing the feedback strength.

\item The relatively steep slope predicted by the weak reionization model
cannot be reduced by decreasing the SN feedback efficiency, i.e., through
decreasing $v_{\rm hot}$ or $\alpha_{\rm hot}$ in the scaling law
(Eq.~\ref{eq:scale_law2}).
\begin{itemize}
\item Reducing $v_{\rm hot}$ preferentially increases the slope of stellar
metallicity -- stellar mass correlation, which can be understood as follows.
Reducing $v_{\rm hot}$ would cause a uniform suppression of the feedback
efficiency expected by the scaling law. Galaxy formation processes can be
affected if the expected feedback strength is lower than the limit set by the
energy condition, which is relatively easier to occur for relatively large
galaxies, e.g., $M_* \sim 10^6\msun$ or higher. So reducing $v_{\rm hot}$ would
preferentially enhance the metallicities of relatively large galaxies, and thus
increase the slope of stellar metallicity -- stellar mass correlation, which is
supported by Figure~\ref{fig:m_z_cor_vhot} (though the results are for the
strong reionization model, for galaxies more massive than $10^6\msun$ the
effects of the reionization are weak).  Our simulation results show that
reducing $v_{\rm hot}$ to $100\kms$ increases the slope from $0.33$ to $0.4$,
and further reducing $v_{\rm hot}$ to $50\kms$ destroys the correlation. 
 
\item Reducing $\alpha_{\rm hot}$ causes a suppression of the SN feedback
efficiency preferentially in small galaxies. This may increase the
metallicities of small galaxies while keep those of relatively large galaxies
unchanged, so it may reduce the slope of the metallicity -- stellar mass
correlation. The lower boundary of $\alpha_{\rm hot}$ is $2.0$, which is set by
the energy condition.  We find that reducing $\alpha_{\rm hot}$ from 3.2 to 2.0
in the weak reionization model does reduce the slope, from $0.45$ to $0.43$;
however, the reduced slope is still too large to be consistent with the
observational one.
\end{itemize}
\end{itemize}
Hence the observed slope of the stellar mass -- metallicity correlation
strongly prefers the strong reionization model.  As mentioned in
Section~\ref{sec:model}, the weak reionization model is close to or slightly
later than the cosmic average reionization epoch and the strong reionization
model is probably earlier than the cosmic average.  The strong reionization
model implies that the region around the MW or the local group is reionized
earlier than the cosmic average, or there is also a non-negligible contribution
from the local ionizing radiation field in additional to the global ionizing
background.  This is in agreement with the conclusion drawn in
\citet{font2011}. Our results support the patchy reionization scenario
\citep{L12}, in which the reionization is inhomogeneous and the reionization
redshifts for different halos are different.  Note that our work tests the
patchy properties on relatively large scales (i.e., the host halo scale $\sim
300\kpc$), that is, the reionization is patchy at least on the scales
comparable to the MW host halo or the local group size.

Apart from the slope, $\sigma (\calZ')/\bar{\calZ'}$ is also mainly affected by
different reionization models (see Fig.~\ref{fig:z_d_reion}). Future
observations on it will further examine the constraint on the reionization
strength.

\subsection{Constraints on the SN feedback models}

After the strong reionization is adopted, the observational slope in the
stellar metallicity versus stellar mass correlation can also further put
constraints on $v_{\rm hot}$.
\begin{itemize}
\item Figure~\ref{fig:m_z_cor_vhot} indicates that only the model with feedback
that is strong enough can provide a relatively small slope. With $\alpha_{\rm
hot}=3.2$, the slope produced by the models with $v_{\rm hot}=100\kms$ or
$v_{\rm hot}=50\kms$ is too large to be consistent with the observation. We
find that even if a larger $\alpha_{\rm hot}$ is adopted, it is still difficult
for the models with a smaller $v_{\rm hot}$ to generate a slope consistent
with observations. For example, if $\alpha_{\rm hot}$ is increased to 4.0, the
slope is reduced only a little to 0.38, which is still larger than the
observational value.  Therefore the observational $\alpha_{\rm obs}=0.30\pm
0.02$ prefers the case of $v_{\rm hot}>100\kms$.

\item The model with $v_{\rm hot}=400\kms$ gives almost the same results as
that adopting $v_{\rm hot}=200\kms$, because the feedback strength is limited
by the energy condition, so these two values cannot be distinguished in the
dwarf satellites.

\item Assuming strong reionization and
$v_{\rm hot}=200\kms$, Figure \ref{fig:m_z_cor_ahot} shows that 
the correlations resulting from the model with  $\alpha_{\rm hot}=3.2$ and that with
$\alpha_{\rm hot}=2.0$ are almost the same. The reason is that the change of the feedback
strength is not very strong.
\end{itemize}

\begin{figure}
\center
\includegraphics[scale=0.55]{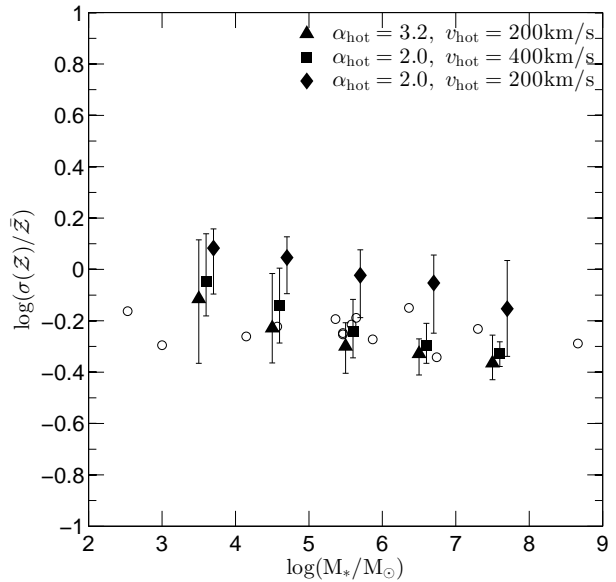}
\caption{Comparison of the observational metal-rich tails with the model
results obtained with different SN feedback parameters. The open circles
represent the observational results, where $\sigma(\calZ)/\bar{\calZ}$ are from
\citet{Leaman2012} and the stellar masses are from McConnachie (2012), for
satellites SMC, Fornax, Sculptor, Sextans, Carina, Leo I, Leo II, CVn I, Seg
I, Ursa Major I, Willman I, Draco, Ursa Minor and Hercules.  The solid symbols
and the error bars have the same meanings as those in Figure~\ref{fig:example},
and the parameters not labeled in the figure have the same settings as those
in fiducial model.  The figure indicates that the fiducial model (solid
triangles) are generally consistent with the observational metal-rich tails
well.
\label{fig:z_d_feedback} } 
\end{figure}

The metal-rich tails are more sensitive to the change in the feedback strength
and may put some further constraints on $(v_{\rm hot},\alpha_{\rm hot})$.
Figure~\ref{fig:z_d_feedback} shows $\sigma (\calZ)/\bar{\calZ}$ obtained from
different SN feedback parameters, together with the observations of $14$
satellites around the Milky Way taken from \cite{Leaman2012}.  The observation
of LMC, with $\log[\sigma (\calZ)/\bar{\calZ}]=-0.33$, is not shown in this
figure, because the stellar mass of LMC is above $10^9\msun$, which is out of
the range of the figure. It is reasonable to omit LMC because here the
discussion focuses on the dwarfs with stellar mass between $10^3\msun$ and
$10^8\msun$.  The observation of another satellite, Bootes I, is not shown in
Figure~\ref{fig:z_d_feedback}, either. The stellar mass of Bootes I is
$2.9\times 10^4\msun$ \citep{local_group2012}, and it has an extremely low
$\sigma(\calZ)/\bar{\calZ}$, $\sim 10^{-3}$, according to \cite{Leaman2012},
which is much lower than the values of all the other observed satellites with
similar stellar masses. This would indicate that Bootes I is very special and
beyond the scope of the model in our work.

Figure~\ref{fig:z_d_feedback} shows some degeneracies in $(v_{\rm
hot},\alpha_{\rm hot})$: the results obtained from the models with
$(400\kms,2.0)$ and $(200\kms,3.2)$ are close, which can be understood from
Figure~\ref{fig:beta_figure} where the feedback efficiencies of the two sets of
parameters are close.  As seen from Figure~\ref{fig:z_d_feedback}, however, the
result of the model with $(200\kms,2.0)$ is distinguishable from the above two.
The observations on $\sigma (\calZ)/\bar{\calZ}$ prefer the models with
$(v_{\rm hot},\alpha_{\rm hot})=(400\kms,2.0)$ and $(200\kms,3.2)$ to the model
with $(200\kms,2.0)$.

\section{Summary} \label{sec:summary}

We have investigated the effects of the SN feedback, the reionization of the
universe, and the molecular hydrogen cooling processes on the chemical
properties of the dwarfs around the Milky Way-like host galaxies, through a
semi-analytical galaxy formation model. Our fiducial model can reproduce the
luminosity function, the stellar metallicity versus stellar mass correlation,
and the distribution of the metal-rich tails of the MW dwarf satellites.

We find that the slope of the stellar metallicity versus stellar mass
correlation is sensitive to the SN feedback parameter $v_{\rm hot}$, which can
lead to a universal change of feedback strength among the whole satellite mass
range.  This slope is larger with smaller $v_{\rm hot}$, but a too small
$v_{\rm hot}$ destroys the correlation. This slope is fairly not changing with
$v_{\rm hot}$ when $v_{\rm hot}$ is above $200\kms$, because the feedback
strength expected by the feedback scaling law is too large so as to be limited
by the energy condition.

We find that the slope of the stellar metallicity versus stellar mass
correlation is also sensitive to the strength of the reionization
process. The slope is flatter if the universe or the local universe is reionized earlier (i.e., reionization is stronger). The reason is that
the halo of a given mass dwarf before infall is relatively more massive in a (local) universe with stronger reionization, which thus allows a more efficient metal enrichment.

Both the SN feedback and the reionization affect the slope, but in different
satellite mass range. Feedback preferentially affects large dwarfs, because the
feedback strength in small dwarfs is limited by the energy condition. The SN
feedback affects the slope through affecting the metallicities of those large
dwarfs, with mass above several $10^6\msun$.  Reionization only strongly
affects small halos, and thus it can only affect small dwarfs significantly. The
reionization affects the slope through affecting the metallicities of the small
dwarfs, with mass below $4\times 10^6\msun$.

We find that the metal-poor tail (quantified by $\sigma (\calZ')/\bar{\calZ'}$)
is sensitive to the reionization. With weaker reionization, the $\sigma
(\calZ')/\bar{\calZ'}$ of the satellites less massive than $10^5\msun$ are
smaller, because weaker reionization increases the star formation rate and
reduces the minor merger possibilities of the small galaxies, which leads to
weaker metal poor tails; but the $\sigma (\calZ')/\bar{\calZ'}$ of more massive
satellites are larger, because weaker reionization lowers stellar metallicities
in these small galaxies, while the metal-poor tails of larger galaxies are
contributed mainly by the small galaxies through mergers, and the reduction of
the stellar metallicities of the small galaxies enhances the metal-poor tails
of the larger galaxies.

We find that the strength of the metal-rich tail (quantified by $\sigma
(\calZ)/\bar{\calZ}$) is sensitive to feedback parameters. With weaker
feedback, the tails are stronger. This is because weaker feedback allows more
efficient metal enrichment and more metal-rich stars to form and thus enhances
the metal-rich tails.

We find that both the metallicity - stellar mass correlation and the
metallicity distribution in individual satellites are not sensitive to
molecular hydrogen cooling, as it only affects the formation of a very small
part of stars in dwarfs.

The various chemical properties can be used to constrain the underlying
physical processes of galaxy formation. The observed slope ($0.30\pm 0.02$) of
the stellar metallicity -- stellar mass correlation prefers the strong
reionization, which suggests that the universe is reionized at a redshift $\ga
10$ or the local universe is reionized earlier than the cosmic average due to
the contribution from the local reionizing sources. This slope also prefers
$v_{\rm hot} > 100\kms$.  The observations on $\sigma(\calZ)/\bar{\calZ}$ can
put constraints on $\alpha_{\rm hot}$: they prefer the models with $(v_{\rm
hot},\alpha_{\rm hot})=(400\kms,2.0)$ and $(200\kms,3.2)$ over the model with
$(200\kms,2.0)$.  There are some degeneracies between $v_{\rm hot}$ and
$\alpha_{\rm hot}$, for example, the results of  $(v_{\rm hot},\alpha_{\rm
hot})=(400\kms,2.0)$ and $(200\kms,3.2)$ are close.
 
\acknowledgements

This research was supported in part by the National Natural Science Foundation
of China under nos.\ 10973001, 11273004, 11373031, and 11390372.  YL and QY
thank the NSF Grant \#1066293, the funds support from the Simons Foundation,
and the Aspen Center for Physics for hospitality, where part of the work was
done. We thank the referee for helpful comments.

\section*{Appendix: Gas cooling recipe}

The gas cooling rate in a DM halo at time $t$, $\dot{M}_{\rm cool}(t)$, is
calculated by
\begin{equation}
\dot{M}_{\rm cool}(t)=\frac{M_{\rm gas}[r_{\rm min}(t+\Delta t)]-M_{\rm
gas}[r_{\rm min}(t)]}{\Delta t},
\end{equation}
where $M_{\rm gas}(r)$ is the total hot gas mass within radius $r$ of the DM
halo and $r_{\rm min}(t)\equiv \min [r_{\rm cool}(t),r_{\rm ff}(t)]$. The
$r_{\rm cool}(t)$ is the cooling radius and obtained by the following
energy-conservation equation in the model of \citet{galform1}:
\begin{equation}
\mu m_{\rm H}\rho_{\rm gas}(r_{\rm cool})\Lambda (T_{\rm gas},Z_{\rm gas})\cdot
(t-t_0)=\frac{3}{2}kT_{\rm gas},
\end{equation}
where $\mu m_{\rm H}$ is the mean molecular mass of the hot gas, $k$ is the
Boltzmann constant, $\rho_{\rm gas}(r)$ is the mass density of the hot gas at
radius $r$, $\Lambda (T_{\rm gas},Z_{\rm gas})$ is the cooling function as a
function of the gas temperature $T_{\rm gas}$ and metallicity $Z_{\rm gas}$ \citep{SD1993},
and $t_{0}$ is the beginning moment of the cooling.  The $r_{\rm ff}$ is the
free-fall radius obtained from the solution of $t_{\rm ff}(r)=t-t_0$, where
$t_0$ is the beginning moment of the cooling process, $t_{\rm ff}(r)$ is the
free-fall time taken by the materials to free fall to the center of the DM halo
from radius $r$.

In the model of \citet{galform1}, at the moment $t_{0}$, the mass density of
the hot gas in a dark matter halo is assumed to follow a distribution with
$\rho_{\rm gas} \propto 1/(r^2+r^2_{\rm core})$, where $r_{\rm core}$ is a
parameter. The $T_{\rm gas}$ is set to be the virial temperature of the DM
halo. The gas heated by the SN feedback (reheated gas) is assumed not to join
the cooling process until the DM halo mass grows up to twice of the mass
obtained at $t_{0}$. Once the DM halo mass doubles, the hot gas distribution is
re-distributed to incorporate the previously reheated gas, and then the
previously reheated gas joins the cooling process.  This treatment partly
realizes the processes in which as the dark matter halo grows, the
gravitational potential changes and the hydrodynamical state of the hot gas
also changes correspondingly.  However, the reheated gas could cool down before
the obvious change of the gravitational potential, which is roughly considered
in the model of \citet{galform3}.  In that model, the reheated gas is assumed
to join the cooling processes in a time scale comparable to the
halo dynamical time scale. In this work, we calculate the cooling of this
reheated gas by the following simple and more detailed treatments.

As the reheated gas is generated at a relatively late time, it is hotter than
the original gas which begins to cool at $t_{0}$, and thus the reheated gas
alone usually would not contribute much to the cooling rate $\dot{M}_{\rm
cool}$. But if the reheated gas is mixed with the cooling hot gas since
$t_{0}$, it can join an efficient cooling immediately.  To calculate the degree
of this mixing process, we assume that the mass density of the reheated gas
follows the same profile shape as the original hot gas.  As this mixing occurs
before the significant change of the halo gravitational potential (i.e., the
halo mass doubles), the parameter $r_{\rm core}$ is assumed not to be changed,
while only the normalization of the distributions are different.  The
temperature of the reheated gas is set to the virial temperature of the DM
halo.  Because of the limited amount of the reheated gas, usually it can only
distribute to an outermost radius $r_{\rm out}$ which is smaller than the DM
halo's virial radius $r_{\rm vir}$. If since time $t'(>t_0)$, $r_{\rm out}$
starts to be larger than $r_{\rm min}(t')$, the reheated gas between $r_{\rm
min}$ and $r_{\rm out}$ is mixed with the original hot gas which cools since
$t_{0}$. The mixing is assumed to be homogeneous and changes the mass density
of the hot gas between $r_{\rm min}$ and $r_{\rm out}$.  We denote the total
gas mass density after the mixing by $\rho'_{\rm gas}$ and the metallicity by
$Z'_{\rm gas}$.  After the mixing, we calculate the cooling radius $r_{\rm
cool}$ from the solution of the following energy-conservation equation:
\begin{eqnarray}
\mu m_{\rm H}\rho_{\rm gas}^2(r)\Lambda (T_{\rm gas},Z_{\rm gas})\cdot
(t'-t_0)+\mu m_{\rm H}{\rho'}^2_{\rm gas}(r) \times \nonumber \\
\Lambda (T'_{\rm gas},Z'_{\rm
gas})\cdot (t-t')=\frac{3}{2}kT_{\rm gas}\rho'_{\rm gas}(r), \nonumber \\
\end{eqnarray}
and the free-fall radius $r_{\rm ff}$ from the solution of the following
equation:
\begin{equation}
\rho'_{\rm gas}(r)=\rho_{\rm gas}(r)\cdot (t'-t_0)/t_{\rm ff}(r)+\rho'_{\rm
gas}\cdot (t-t')/t_{\rm ff}(r).
\end{equation}

\addcontentsline{toc}{section}{REFERENCES}
\bibliography{reference}

\end{document}